\documentclass[sigconf,screen,nonacm]{acmart}

\setcopyright{none}                % removes the copyright notice
\renewcommand\footnotetextcopyrightpermission[1]{} % removes the page-1 box (conference, ISBN, DOI)
\usepackage{amssymb,amsfonts} % Approved with acmart
\usepackage{algorithmic} % Approved with acmart
\usepackage{textcomp} % Approved with acmart
\usepackage{xspace} % Approved with acmart
\usepackage{multirow} % Approved with acmart
\usepackage{colortbl} % Approved with acmart
\usepackage{subcaption} % Approved with acmart
\usepackage{soul} % Approved with acmart
\def\BibTeX{{\rm B\kern-.05em{\sc i\kern-.025em b}\kern-.08em
    T\kern-.1667em\lower.7ex\hbox{E}\kern-.125emX}}

\makeatother

\begin{document}

% "Probabilistic" instead of "Generative"?
% IR-based Autotuning in Multistep Regime
% Accelerating Autotuning with IR-Based Analysis
% Transfer-Learning-Guided IR Selection for Autotuning
% Accelerating Transfer-Learning Autotuning with Predictive IR Performance Ranking
\title{Accelerating Transfer-Learning-Based Autotuning with Predictive LLVM IR Performance Ranking}
%\subtitle{\normalsize{SC 2025 Submission \textbf{\#NaN} -- Confidential Draft -- Do NOT Distribute!!}}

% EDITOR COMMANDS:
% Comment this out to delete all editor comments
\newcommand{\editorComment}[1]{%
%#1
}

\newcommand{\tlr}[1]{%
\editorComment{ {\color{blue}TR: #1} }
}
\newcommand{\arafat}[1]{%
\editorComment{ {\color{orange}Arafat: #1} }
}
\newcommand{\akash}[1]{%
\editorComment{ {\color{red}Akash: #1} }
}
\newcommand{\rge}[1]{%
\editorComment{ {\color{green}RG: #1} }
}
\newcommand{\ali}[1]{%
\editorComment{ {\color{green}AJ: #1} }
}
\newcommand{\xingfu}[1]{%
\editorComment{ {\color{green}XW: #1} }
}

% MUST BE IN DOUBLE-BLIND FOR SUBMISSION TO SC
% ENSURE THIS MATCHES THE ACM TEMPLATE: https://hpcrl.github.io/ICS2025-webpage/for-authors/ics2025template.zip
% PAGE LIMIT: 10 PAGES W/O REFERENCES
%\author{\IEEEauthorblockN{Anonymous authors}}
% \author{\normalsize{SC 2025 Submission
% \textbf{\#NaN} -- Confidential Draft -- Do NOT Distribute!!}}

\author{Md Arafat Hossain}
\authornote{Both authors contributed equally.}
\affiliation{%
  \institution{Department of Computer Science}
  \institution{Iowa State University}
  \city{Ames}
  \state{Iowa}
  \country{USA}
}
\email{arafat@iastate.edu}

\author{Thomas Randall}
\authornotemark[1]
\affiliation{%
  \institution{School of Computing}
  \institution{Clemson University}
  \city{Clemson}
  \state{South Carolina}
  \country{USA}
}
\email{tlranda@clemson.edu}

\author{Akash Dutta}
\affiliation{%
  \institution{AMD}
  \city{Ames}
  \state{Iowa}
  \country{USA}
}
\email{Akash.Dutta@amd.com}

\author{Xingfu Wu}
\affiliation{%
  \institution{MCS}
  \institution{Argonne National Laboratory}
  \city{Lemont}
  \state{Illinois}
  \country{USA}
}
\email{xingfu.wu@anl.gov}

\author{Rong Ge}
\affiliation{%
  \institution{School of Computing}
  \institution{Clemson University}
  \city{Clemson}
  \state{South Carolina}
  \country{USA}
}
\email{rge@clemson.edu}

\author{Ali Jannesari}
\affiliation{%
  \institution{Department of Computer Science}
  \institution{Iowa State University}
  \city{Ames}
  \state{Iowa}
  \country{USA}
}
\email{jannesar@iastate.edu}

%\renewcommand{\shortauthors}{}

% ACMART requires abstract/keywords here

\begin{abstract}
As the complexity of High Performance Computing (HPC) ecosystems continually increases, achieving optimal performance becomes a challenge.
Traditional performance autotuning techniques provide promising means to navigate this complexity, these techniques remain computationally intensive and require many evaluations to find optimal configurations.
%They are often limited by an inability to reuse knowledge or generalize too broadly to identify the absolute best performance.

%Transfer learning-based autotuning methods address some of these challenges by leveraging prior knowledge, but often fail to improve over longer search durations or require extensive calibration before improving within new domains.
%These limitations create a pressing need for a data-efficient and intelligent autotuning mechanism that can rapidly identify optimal configurations with minimal computational cost while supporting longer-term searches to refine performance.

This work proposes an autotuning framework that designs a machine learning-based ensemble LLVM Intermediate Representation (IR) ranker, Neural Configuration Scorer (NCS).
NCS ranks the performance of IRs sampled by a transfer-learning-based autotuner, improving the efficiency of the tuning process by reducing tuning overheads and circumventing subpar evaluations.
By leveraging knowledge from related tasks, we are able to effectively exploit the transfer relationship to access high-performing configurations in fewer samples than traditional techniques that rely upon iterative refinement.
% Although this study leverages the benefits of transfer learning, its combination with an intelligent ranker, allows it easily order configurations in terms of performance, which prior transfer learners struggle to do.
% In this work, we propose Neural Configuration Scorer, an ensemble framework that ranks the performance of LLVM Intermeidate Representations (IRs) for transfer learning-based autotuning.
% Rather than predicting the performance of all program variants, we focus on relatively ranking a near-optimal subset of IRs produced by prior transfer learning techniques.
%Prior work provides a probabilistic method for transfer learning, which we use as a supporting model to limit the number of IRs that are considered and evaluated.
%incorporate the Gaussian Copula for Transfer Learning Autotuning (GCTLA) with a convolutional neural network (CNN)-based ranking mechanism, namely Neural Configuration Scorer (NCS) to develop a novel autotuning technique that enhances generative autotuning.
%Initially, we leverage GCTLA as a supporting model to generate many near-optimal configurations in the search space.
%Then we utilize NCS to rank the expected performance based on the Intermediate Representations of configured programs.
%Finally, we evaluate the best candidates to determine the actual peak performance.
%This combination significantly reduces the need for exhaustive empirical evaluations while maintaining high accuracy in identifying optimal configurations.
Our framework can %expand GCTLA to longer-form searches and
achieve similar performance improvements as state-of-the-art autotuning techniques with up to 61.67\% fewer evaluations, averaging 27.85\% fewer evaluations across various HPC benchmarks.

\end{abstract}

%\begin{IEEEkeywords}
%\end{IEEEkeywords}

\maketitle
\pagestyle{plain}
% GENERAL DRAFT THOUGHTS:
\begin{comment}
Related works should be placed BEFORE our design (immediately after background) to discuss the gap in the literature that we see and situate our solution.
Experiments need to clearly show the end-to-end methodology; this may be elaborate enough to require a standalone section prior to the results.
\end{comment}

\section{Introduction} \label{sec:intro} The complexity and scale of High Performance Computing (HPC) ecosystems have exploded over the past decade \cite{GoldenAgeArchitecture}.
Tuning and optimization of applications on HPC systems
% Application tuning and optimization on High Performance Computing (HPC) systems 
are worthy uses of compute time and resources; however, pursuing near-optimal configurations remains prohibitively very time-consuming.
% Existing literature provides sophisticated means for learning performance relationships from scratch or with the aid of related information, but requires either large quantities of information that are costly to obtain or necessitates suboptimal explorations of the search space to reduce the likelihood of being trapped in local minima.
Existing literature provides sophisticated means for modeling performance or directly simulating hardware.
However, these approaches generally require either large quantities of data that is costly to collect or limit cost at the expense of precision and flexibility, providing sub-optimal results.

% We combine the Multimodal GNN Autotuner (MGA) \cite{MGA} and Gaussian Copula for Transfer Learning Autotuning (GC\_TLA) \cite{GC_TLA} to create a new autotuning approach

Performance autotuning relies upon discriminating between highly similar program configurations to determine the optimal configuration with respect to one or more objectives such as runtime, energy efficiency and throughput. In this paper, 
we focus on optimizing the runtime of semantically equivalent source-code expressions of programs with respect to the programs' input sizes.
Predicting these figures of merit using IRs remains challenging, but IR-based performance predictions have largely been limited to generalized program runtime or estimates of the cycle count occupied by small code segments~\cite{DiffTune,DiffTuneRevisited}.
Despite this prior research and other rich opportunities within IR modeling, performance autotuning has traditionally favored costly empirical performance observations rather than IR-based performance inference.
The primary limiting factor has been the massive overhead of generating thousands to millions of IRs that span the entire search space.

In this work, we propose a novel approach which combines a new machine learning-based IR ranking model with GCTLA~\cite{GC_TLA}, an existing method for probabilistic transfer learning, to cleverly limit the number of IRs under consideration.
The combination is able to greatly improve transfer-learning-based autotuning efficiency for new tasks --- i.e., unseen input sizes for a program executed on a particular system.

Our key contributions in this paper include:
\begin{itemize}
    \item %Framework for using IRs in the autotuning process
    % We provide a framework to effectively leverage CNN ensembles within the domain of transfer-learning-based performance autotuning.
    Proposing a framework for transfer-learning-based autotuning with the LLVM IRs of the programs.
    \item %Model using IR
    % Our technique for leveraging CNN ensembles, 
    Designing a machine-learning ensemble, NCS (Neural Configuration Scorer), that predicts performance from LLVM IRs, granting it higher fidelity to actual hardware behaviors than the abstract high-level source code changes that form the tuning space.
    \item % Improvement using few-shot transfer learning
    Reducing the number of evaluations required for few-shot transfer learning with minor overhead.
    We demonstrate that the combination of GCTLA+NCS is mutually beneficial and further demonstrate the reliability of NCS through the lens of post-hoc analyses of combining state-of-the-art iterative techniques with NCS.
    \item Our performance evaluations in few-shot transfer-learning settings require an average of 27.85\% fewer evaluations to reach optimal results than prior works, with up to 61.67\% fewer evaluations.
\end{itemize}
    %\item We propose an IR-based neural ranking model (NCS) that permits generative autotuning to perform long-term searches where it was previously only usable for few-shot evaluations.
    %\tlr{ Omit : \item By coupling GCTLA with CNN-based performance estimation, our method leverages the strengths of both generative and discriminative models, framework for modern HPC applications.}
    %\item We permit the use of IRs as performance indicators to prioritize
    %\item We improve the few-shot effectiveness of GCTLA \textcolor{red}{(first time we are mentioning this. We should provide some context before)} as a supporting model for NCS, requiring an average of 83.56 fewer evaluations to identify the search optima.
    %\tlr{Conclusion notice only: \item We create a large-scale open-source dataset of empirical autotuning information for use by the research community.}

%The regressive performance prediction model, built using a CNN, evaluates generated configurations and prioritizes those most likely to yield optimal performance. Unlike traditional predictive models that require extensive training data, our CNN-based model learns to distinguish high-performing configurations within a narrow search space, making it significantly more efficient.

\begin{comment}
Our code and tuning data are publicly available at this URL: \url{}.
\end{comment}
%We intend to provide open-source versions of our code and all related tuning data after the blind review.

The remainder of this paper is organized as follows. Section~\ref{sec:background} provides key background information.
Section~\ref{sec:challenges} discusses promises and challenges.
Section~\ref{sec:design} proposes the framework design in detail.
Section~\ref{sec:experiments} presents the performance evaluation.
Section~\ref{sec:related} describes the related work.
Section~\ref{sec:conclusions} concludes the paper.
\section{Background} \label{sec:background} % Should introduce concepts of:
% * Search surrogates
% * Ways to represent surrogates (DNN based on IR, GP, multi-arm bandits)
% * Ways to reduce search space size (GC, RF, multi-fidelity)
% * Cost of tuning and why we prefer few-shot transfer tuning as much as possible
% * Utility of tuning (when is cost worth it)

% \rge{Overlaps between background / related works, can be made simpler. Move related works or challenges as needed. \xingfu{Could combine background \& related works}}

\subsection{Performance Autotuning}

% Search spaces and objectives
Performance Autotuning~\cite{wu2022autotuning, balaprakash2018autotuning, behzad2019optimizing, wu2023ytopt}, also known as autotuning, is a class of algorithmic techniques that helps improve the performance of HPC applications/libraries via a set of tunable parameters.
% optimize the performance of a kernel for a given platform via a set of tunable parameters.
Speedup, throughput, energy usage and latency are typical performance objectives in autotuning, which may be tuned singularly or in combination by algorithms that support multi-objective optimization.
Tuning parameters are typically expressed as code-level optimizations, compiler options and runtime settings. %such as loop tile sizes, memory management strategies, or hardware-specific tuning e.g., CPU affinity, cache usage, etc.
% For example, in matrix multiplication kernels, autotuning can dynamically adjust loop transformations ~\cite{kruse2020autotuning}, memory layouts~\cite{youssef2022autopager}, thread parallelization~\cite{popov2019efficient} strategies to improve computational efficiency.

%Due to the vastness of parameter spaces and cost of empirically determining performance objectives, autotuning requires  extreme sampling efficiency.

% MIA but maybe here: surrogates
% Ways to reduce search space
The primary challenges of HPC autotuning are induced by the size of search spaces (often defined in a combinatorial manner, growing exponentially), the nonlinear complexity of the relationship between configurations and performance, and the high cost of observing the performance relationship.
Various autotuning techniques have to strike a balance between three key characteristics:

\begin{itemize}
    \item Training and refitting cost: The computational burden associated with training, fine-tuning or utilizing the autotuning model beyond the empirical evaluation cost.
    \item Immediate efficacy: The maximum quality of results achieved within the first few evaluations.
    \item Ultimate efficacy: The maximum quality of results achieved by the end of a search's full term. For techniques that can try all possible combinations, ultimate efficacy may be better measured when the global optimum is selected.
\end{itemize}

Excelling at any pair of characteristics requires sacrificing the third.
For instance, a technique that emphasizes ultimate efficacy and lower training cost must necessarily have poor immediate efficacy as it cannot know what to avoid in the near-term.
%Any well-motivated technique that emphasizes only one of immediate or ultimate efficacy will further excel at its task with greater training and refitting cost.

Efficient tuning techniques aim to limit the number of empirical observations while effectively modeling the complex relationship between the objective function and the parameter search space.
% Efficient tuning algorithms typically focus on reducing the number of observations required to model the complex relationship as accurately as possible so that the search space does not need to be extensively explored, achieving a high degree of ultimate efficacy within a limited number of empirical observations or within a given time limit.
% In order to accurately model the entire performance relationship as quickly as possible, some sub-optimal observations are necessary.
Sub-optimal evaluations are common during the tuning process, but should be limited as the global best and global worst are frequently separated by an order of magnitude.
% The sub-optimal evaluations can be extremely costly, especially when determining runtime performance, as the global best- and global worst- runtimes are frequently separated by an entire order of magnitude.
%Given the nonlinear and complex relationship between the parameter configurations and performance outcomes, autotuning aims to identify optimal configurations without measuring performance objectives for every possible configuration in the parameter space.
HPC autotuning is made even more challenging as the optimal configuration often varies based on the input size and the target architecture.
Overcoming this challenge manually is time-consuming and costly, which leads to the necessity of intelligent search strategies, such as machine learning-based or evolutionary algorithm-based strategies to navigate the vast search space and map the relationship among different parameters.
% Autotuning is especially imperative for large-scale HPC applications, as they are resource-extensive and need tuning at the same time.
%The number of possible configurations is vast.

Techniques including GCTLA~\cite{GC_TLA}, Random Forests (RF)~\cite{bei2015rfhoc}, and multi-fidelity~\cite{kandasamy2017multi, zhu2022gptuneband, kandasamy2016gaussian} are very efficient in reducing the search space by focusing on probable high-performance regions.
Such approaches have proven to yield a substantial amount of performance gains~\cite{balaprakash2018autotuning, falch2017machine}.
By automating the search for optimal settings, autotuning helps HPC applications enhance their performance without manual intervention, making it a pivotal tool for the maximization of the capabilities of modern supercomputing systems. 

% Not sure this truly requires a background section -- we are not doing anything novel to the IR itself, just ingesting it.
\subsection{Intermediate Representation}
%how objective time informations are captured in irs
LLVM (Low-Level Virtual Machine) IRs (Intermediate Representations) serve as an abstraction of source code that bridges high-level languages with machine code.
They allow compilers to perform optimizations and transformations on source code in a consistent manner, regardless of the original language.
The IR is a desirable point for performance analysis due to its simplicity, expressiveness, and abstraction.
IRs are hardware independent and enable performance analysis without being tied to a specific source language. 
% or compiler flags.
They also contain semantic and structural information about the source code, including, but not limited to data flow, control flow, memory access patterns, loop insights, store and load instruction details etc. crucial informations that are essential in detecting performance patterns, making them ideal for machine learning based feature extraction and performance prediction tasks.

% We focus on the LLVM IR 
% (Low-Level Virtual Machine Intermediate Representation), 
% developed as a part of LLVM~\cite{lattner2004llvm} 
LLVM IRs are a community standard for code representation for optimization and tuning tasks~\cite{venkatakeerthy2020ir2vec, cummins2021programl, dutta2024mirencoder} and generating~\cite{kalms2018automatic} code in a variety of languages.
The LLVM IR still retains several key markers that assist in performance analysis, such as mangled function names that permit restricting analysis' scope to only the code regions that are affected by optimization efforts.
Unlike the LLVM's Machine Code Analyzer, which estimates the cycle counts and IPC of straight-line assembly fragments, the LLVM IR still expresses loop behaviors that compose many useful optimizations such as unrolling, tiling and loop interchanges.

\subsection{Transfer Learning in Autotuning}
Transfer learning improves the efficiency of performance of a target task by leveraging the knowledge from previously learned related tasks \cite{5288526}, which is crucial for machine learning models.
Generally, a number of ``source'' tasks are given as prior knowledge, representing the performance relationship in at least two settings so that differences in performance with respect to tasks can be observed.
Then, a ``target'' task is proposed, wherein the transfer model must accurately adjust insights gained from the source tasks and their relationship to one another to predict the relationship that will be represented in the target domain.
% Transfer learning enables a machine learning model to generalize knowledge between related tasks.~\tlr{CITATION NEEDED}.
This is highly desirable in HPC autotuning, where the performance relationship between various architectures or inputs will change but should maintain key traits that may have been learned via previous tuning efforts.
Efficient reuse of prior tuning information can permit faster approximations of the optimum on new tasks and aids in avoiding regions of the space that are likely to be very costly to evaluate.

Transfer learning is typically evaluated in a few-shot context, where the transfer model must immediately apply its knowledge to maximal effect in a small number of evaluations~\cite{GC_TLA}.
However, other transfer learning techniques utilize the transfer to get a head start on the autotuning relationship and iteratively refine the new task relationship as data is collected~\cite{zhu2022gptuneband}.

%In this paper, we combine transfer learning based autotuner with a neural-network based ranking tool to provide robust configuration prediction for HPC applications. \tlr{``In this paper, ...'' is usually not background material}
% Needs to be expanded to discuss the tradeoffs between different approaches and resource-budgets (larger budget = able to explore more, smaller budget = need to exploit effectively with available knowledge)

\section{Promises and Challenges} \label{sec:challenges} \tlr{Classification of the parameter space based on objective values is an ``easier'' problem than regressing the objective value from parameters / IR (whether via CNN or GNN).
The parameter space is the simplest/most direct representation, some division there DOES divide performance.

However, that's just repeating the GC's job and we know it's hard to do that job better via IRs because you have to make ALL of the IRs -- we're working within a subset provided by GC to sidestep that problem.
The changes that improve performance should be reflected in IRs, but grouping by performance may combine MULTIPLE parameter sub-groups together, so no guarantee that the performance objective neatly organizes the search space.

* This is what makes classification difficult -- figuring out parameters from the IR is quite easy, but figuring out performance variability within classes is quite hard

* We can show via importance modeling on the data we have (for 3mm -- only one I've checked so far) that the parameter importance is mostly static WRT problem size and performance class within size. So the IMPORTANCE doesn't change, but the important values DO CHANGE.

* Noise in the dataset means predicting a single class most of the time leads to lower error

* Relaxation of the problem for CNN by splitting problem regions between models, GNN is larger and more advanced / closer to actual representation, should navigate this more easily
}

Many autotuning techniques~\cite{wu2023ytopt,GPTuneBand,OpenTuner,bliss} perform iterative refinement over surrogate models.
% Essentially, these techniques iteratively propose new samples in the search space, evaluate them, and update their knowledge of the performance-configuration relationship across the space.
These models maximize sample efficiency, a key trait for reducing the cost of autotuning.
However, this requires serially evaluating singular points within the search space, often bottlenecking the learning process on performance observations.
Furthermore, a significant number of observations are needed for sophisticated techniques to hone in on the optimum, meaning that the cost of autotuning remains prohibitively high despite the extreme sample efficiency.
%To identify the global optimum for new tasks, they require a substantial number of samples that provide good coverage of the entire search space, making them unsuitable for few-shot strategies.

Evaluating the potential performance of a source code reconfiguration via the program's LLVM IR is often significantly cheaper than a complete evaluation, however prior techniques have not leveraged IRs because the search spaces are too large to comparatively evaluate all possible variations within common search spaces.
% This requirement makes them unsuitable for few-sample strategies, also referred to as few-shot strategies, in new task scenarios.
However a recent transfer-learning autotuner, GCTLA~\cite{GC_TLA}, achieves comparable quality few-shot results without relying on iterative search.
The technique forms a probability model of transfer-learning autotuning to permit sampling many high-performing candidate configurations for new tasks without prior iterations or empirical evaluations in the new domain.
The sampled subset includes at least one near-optimal configuration with a high level of confidence.
%By empirically evaluating the configurations from this subset, GCTLA typically identifies the best-performing configurations within several attempts.

Despite the promise of identifying high-performing configurations within a few shots, current probabilistic autotuning via GCTLA~\cite{GC_TLA} suffers from two major challenges:
%perform well in few-shot transfer learning settings, but do not necessarily scale well to utilize the available resources typically allocated for tuning.
%The primary limitations of using the GCTLA model beyond few-shot transfer learning are as follows:

\begin{itemize}
    \item \textbf{Non-Discriminative Nature of the Sampled Subset.} GCTLA %have a high level of confidence that the sampled subset contains some of the best-performing configurations; however, we
    lacks a reliable method to determine which
    %ones
    sampled configurations are more promising than others.
    A straightforward approach to identifying the best candidate is to empirically evaluate each one.
    While this can be less expensive than iterative approaches, it is not ideal for longer-duration tuning.
    %The generative model has no reliable discriminator to prefer one generated parameterization over another, meaning that all generations have to be evaluated to determine the best candidate.
    \item \textbf{Degrading into Random Search.} %\textbf{Decreasing Proportion of Best-Performing Configurations as the Generated Subset Size Increases.}
    The only way to identify additional high-performing configurations is to increase the sample size. % we need to increase the size of the generated subset
    However, as the sample size increases, the proportion of best-performing configurations within it decreases, gradually degrading towards a random sample of the overall population.
    Consequently, the potential performance gains become disproportionate to the evaluation cost when using a straightforward approach.
    % increases but the  increase, leading to more evaluations  Continually using the model to propose new configurations will ultimately degrade into random sampling, meaning that longer-term searches are inappropriate.
\end{itemize}

%Exceeding the predictive evaluation budget has rapidly diminishing returns that are exaggerated by the lack of mechanism to reject a proportion of the population when over-sampling.

%Generative techniques such as GCTLA minimize the training data cost and provide quick access to high-performing areas of the search space. While computationally cheap, the technique does not consistently find the absolute best performance and gradually degrades into a randomly ordered traversal of the search space as additional evaluations are collected. The limited number of useful evaluations limits GCTLA's practical utility for most autotuning use cases. To support longer searches while remaining competitive with prior art, insightful methods to overcome these primary limitations are required.

% Design should point out that not all techniques are anti-thetical to one another; utilizing increased training cost can permit two techniques to collaborate on immediate- and ultimate- efficacy
% The trick is to minimize the divergence in prerequisites to align the training costs
% We may also discuss why the other combination (Ultimate->Immediate) does not generally work: Ultimate-term efficacy requires advanced understandings that immediate-term efficacy models cannot adequately provide

The limitations of GCTLA can be addressed by accurately predicting the performance of samples at lower cost than empirical evaluation, which can be performed by analyzing the program's LLVM IR.
We propose a machine learning-based performance model that can rank the sampled IRs.
Because our approach relies upon IRs, there is no need to calibrate performance between related tasks and training is only required to fit the models to wholly new domains (ie: new applications).
%Similar to generative techniques, our proposed predictive models are trained on configurations from previous tasks and do not require empirical sampling or evaluation of configurations from new tasks.
Together, the combined approach of probabilistic sampling and predictive performance models maintains low cost, making it well-suited for few-shot performance autotuning.

% Our work proposes utilizing LLVM IR performance prediction as a low-cost means to rank generated samples from GCTLA.
\tlr{An easy challenge we could mention is the Halting Problem -- you cannot predict performance absolutely without actually executing the code, so anything we can do faster than execution is necessarily inaccurate to some degree.}
% Predicting performance from IRs carries lower cost than empirically observing performance.
% However, these predictions are still challenging --- small changes in the source code can dramatically alter many disparate regions of an IR as it is a lower-level representation.
\tlr{Grouping based on performance may combine many different IRs together, which is difficult to learn a relationship on.
Combining similar IRs with respect to the search space does not help with discriminating performance.
We also have limited data that we could point to that demonstrates the importance of parameters don't change a lot but their optimal values do, meaning that picking the best IR is a very nuanced operation.
}

%Predicting IR performance trades off incremental knowledge gains about all levels of performance in the tuning space for faster access to the high-performing area of interest.
%Our framework leverages IRs as a cost-effective proxy for evaluating and ranking performance. % generated code samples produced by GCTLA.
% This approach offers a practical alternative to direct execution, which is often computationally expensive.
Performance prediction is a complex challenge, particularly within the context of HPC applications, where execution behavior exhibits significant variability across different hardware architectures, optimization strategies, and runtime configurations.
Autotuning represents an especially challenging domain for performance prediction, wherein training data is limited and programs with only minor differences yield highly varied performance.
\tlr{Over-generalizing models will predict a single class or a simple performance mean to reduce error, but this also harms the ability to differentiate performance!}
%Unlike iterative approaches that benefit from increased knowledge over time, our IR performance predictions can only serve to prioritize evaluations training a predictive model for a generative approach is more challenging as it shifts the thrust from iterative sampling to a single-shot generation of high-performing configurations, requiring an upfront understanding of IR and performance dynamics.   

\tlr{This needs to blend in better -- the shift is too dramatic. Is it just missing a topic sentence for the paragraph to help the reader along?}

% Contrastive learning with triplet loss
% \tlr{CITATION NEEDED -- readers might not know it} or transformer-based approaches \tlr{citation requested -- is there something we can point readers to?} struggles to generalize the relationship between IR and performance (execution time) in a few-shot setting as they require large amounts of data to learn meaningful differences in IR structures and execution contexts.
% Coupled with the substantial amount of time required by these approaches to train from scratch or to finetune a pretrained model, they become impractical for usage in HPC autotuning where data scarcity is a primary limitation.
% Traditional ensemble learning approaches fail as well for performance prediction due to the large search space and the hierarchical dependencies within LLVM IR, leading to suboptimal results limiting their utility in autotuning tasks.
% \tlr{I think that I follow this due to familiarity with the work, but we need to revise it so a new reader can follow it easily.}

\tlr{Other challenges that we encountered as we developed this work, may or may not bear mentioning:
\begin{itemize}
    \item GC is not advanced enough to handle the complexity of embeddings (attempt to train GC on MGA embeddings to predict an embedding that MGA may decode into a nearest-neighbor IR search for a reversed relationship)
    \item Using K-Nearest Neighbors proved difficult due to class imbalance and minimal training data (even when inverted train/test, class imbalance and variability between classes proved very troublesome)
    \item Using DNN / LLM -ish techniques for predicting IR performance proved to be very difficult (mapping the scope of changes from source code to IR is not straightforward, the state space is difficult to interpret and IRs get very large on certain benchmarks which challenges these techniques)
    \item Identifying a useful problem for GC + IR-based techniques was somewhat difficult. It also puts a big resource cost on producing results.
\end{itemize}

We also had some prior analyses that may hint at challenges existing, but we probably won't use them in the paper:
\begin{itemize}
    \item Due to the pidgeonhole principle, there isn't a chain of 1-element edits in the configuration space that easily explains search viability (a notion of complexity taken from exhaustive syr2k data)
    \item Similarly, solving the ``best training dataset'' to produce the global optimum is also very challenging due to the curse of dimensionality
    \item Sensitivity analyses on syr2k exhaustive data can indicate that the importance of variables changes across the transfer domain as expected, but in quantitative form.
\end{itemize}
}
\section{Design} \label{sec:design} % Should introduce concepts of:
% * Initial training dataset collection (ie: BO via ytopt)
% * GC as a tool for sampling near-optimal selection of search space (generative autotuner)
% * DNN as a tool for discriminating performance based on IRs

\tlr{Continue to keep perspective on diminishing returns of GCTLA here}

\begin{figure*}[h]
    \centering
    \includegraphics[width=0.9\linewidth]{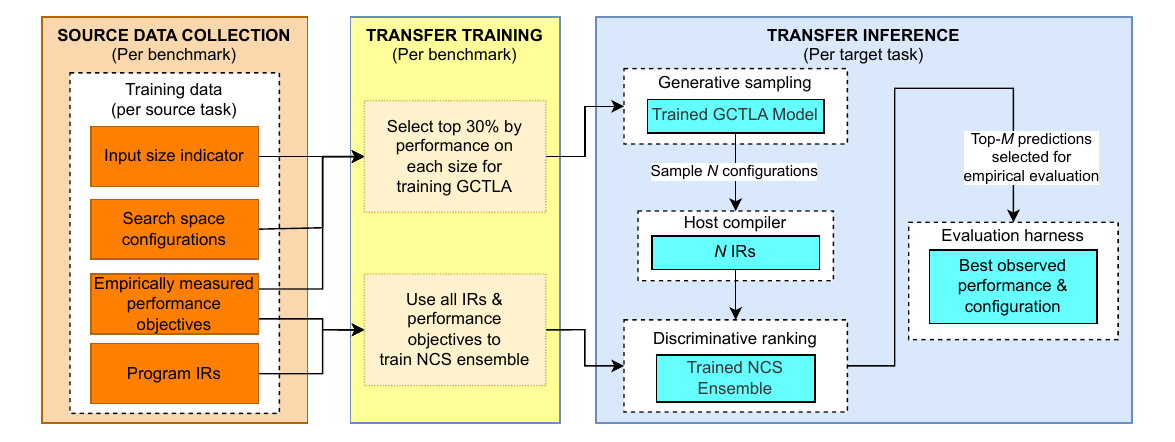}
    \caption{Our approach uses IRs and configurations from source task data to fit both GCTLA and NCS models. The GCTLA model generates high-performing configurations on a transferred task, which are ranked by the NCS ensemble. The highest ranked generations are empirically evaluated to determine the best-performing configuration.}
    \label{fig:overall_design}
\end{figure*}

In this section, we present our novel combination of the probabilistic transfer-learning based autotuner and neural ranking model along with special considerations made for these components in isolation as well as in combination.

The overall design is presented in Figure~\ref{fig:overall_design}.
For a given benchmark, at least two source tasks must be represented.
Each source task is composed of four components: a task indicator, evaluated configurations from the search space on the source task, and the corresponding IRs and empirically measured performance objectives.
%We only use configurations and associated IRs from source tasks to train the GCTLA and NCS models.
The training data is further described in Section~\ref{sec:training_data_description}, with additional minutiae about our experimental training settings in sections~\ref{sec:train_GCTLA} and~\ref{sec:train_NCS}.
Once trained, the GCTLA model samples a number of configurations and we create the corresponding IR for ranking with our Neural Configuration Scorer (NCS) model.
The NCS model allows us to sort all sampled IRs from best-to-worst expected performance; the highest-ranked IRs will be lowered to machine-specific code and evaluated to determine the highest-performing configuration.

\subsection{Probabilistic Autotuner} \label{sec:train_GCTLA}

Despite the limitations of GCTLA noted in Section~\ref{sec:challenges}, training a probabilistic transfer model requires minimal data and only requires data from source tasks.
Additionally, sampling from the learned probability distribution is computationally negligible, meaning that extra time is available to produce the IRs for use with NCS.

As such, we follow the recommendations from~\cite{GC_TLA} to build our supporting model.
This includes the procedure for forming training datasets for transfer-learning and using the top-30\% of source task data to fit the GCTLA model.
However, instead of sampling a number of configurations based on the budgeted expectation, we intentionally over-sample to create a larger pool of candidates.
Then, we use NCS as a discriminative process to sort the sampled candidates and prioritize the highest performing configurations.
This method allows us to alleviate a primary limitation of GCTLA and extend its utility to longer-duration searches where it would normally be subject to diminishing returns, provided that the following conditions can be satisfied.
First, comparisons should carry low cost to prevent overheads from overshadowing improvements to search efficiency.
Second, the comparisons should be as precise as possible near the optimum to observe the best candidates as early as possible.

The latter condition represents a novel opportunity for autotuning and a key contribution of our work.
Typically, success in autotuning requires high-fidelity modeling across the entire space to properly disambiguate all levels of performance.
Because the samples from GCTLA provide cheap means to cull low-performing areas of the search space from consideration, the NCS model only needs to accurately discriminate between the near-best configurations in the space.
This significantly reduces complexity and the amount of data required for NCS to succeed.

\subsection{Neural Configuration Scorer (NCS)} \label{sec:train_NCS} \label{sec:model_architecture}

NCS is the key component of our efficient autotuning pipeline.
This section outlines the relevant design choices and how they impact the autotuning process. 
NCS can be very briefly described as an ensemble model that comparatively ranks the performance of near-optimal IRs.
But as mentioned in Section \ref{sec:challenges}, several challenges must be addressed first.

% original text as of 9.49PM 04/14 -> We construct a model that ranks IR performance amongst near-optimal IRs.
% But as mentioned in section \ref{sec:challenges}, there several challenges that complicate performance prediction. 

\begin{figure}[t]
    \centering
    \includegraphics[width=0.7\linewidth]{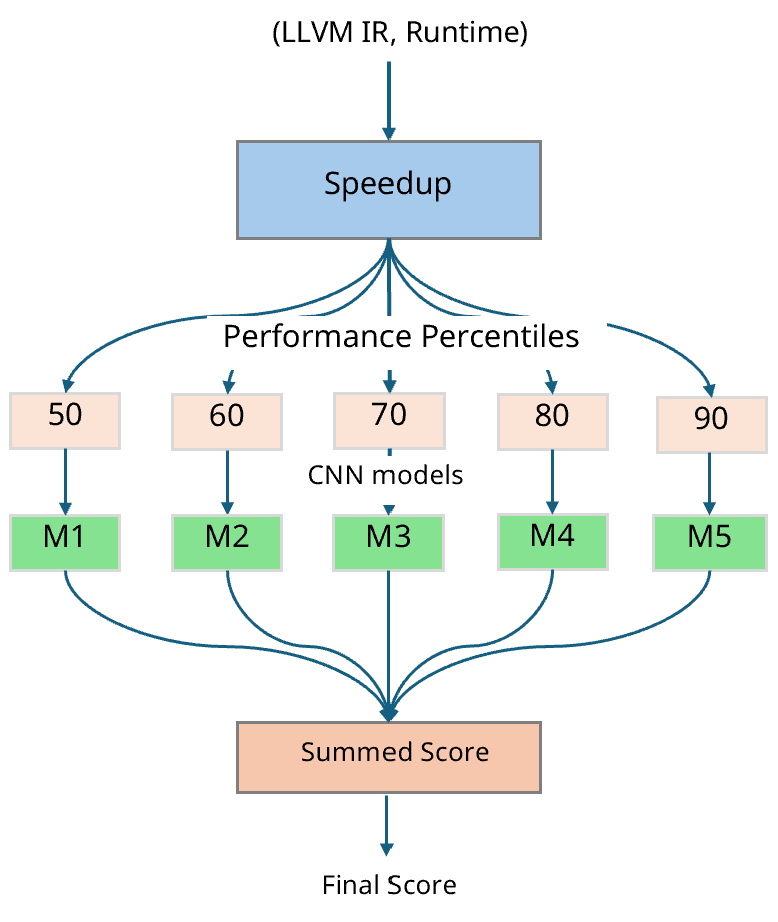}
    % \vspace{-2em}
    \caption{NCS intakes IRs and their corresponding runtimes. Each IR is tokenized, and each runtime is normalized to standardize the data. Next, the data is divided by performance percentiles (50-60\%, 60-70\%, ... , 90-100\%), each of which is used to train a dedicated CNN model that classifies if the IR's performance will meet or exceed its target range. Finally, the predictions from each CNN model are summed to produce a comprehensive final score for ranking the configurations.}
    \label{fig:ncs_overview}
\end{figure}

% an extremely challenging task in HPC environment.
%Keeping that in mind,
% Creating the IR of a program is often less costly than directly evaluating a program's runtime, but compiling all possible IRs across the entire search space is overly burdensome.
%\tlr{We might not introduce it as challenging -- Additionally, as outlined in Section~\ref{sec:challenges}, accurately predicting performance objectives such as runtime from the IR is immensely challenging.}
\tlr{Arafat began using ChatGPT o1 here}
Unlike conventional machine learning approaches that focus on single-class classification or discrete label prediction, we aim to produce a continuous scoring function that reflects the relative performance potential of each generated sample.
Instead of predicting a category of performance, we assign scores such that higher scores will correspond to samples with superior performance.
To this end, we utilize an ensemble of convolutional neural networks, each independently evaluating the LLVM IR representations and assigning an unweighted score to each.
Each network within the ensemble targets a specific performance percentile of the dataset, ensuring specialization in different performance ranges. % and improving the overall ranking accuracy.
The final performance score for a given sample is obtained by the summation of individual scores from all models in the ensemble as shown in Figure \ref{fig:ncs_overview}.
As such, if all models agree that an IR exceeds their target threshold of performance, the IR is ranked highest, with each doubtful member of the IR equally lowering the score.
%This scoring mechanism facilitates the ranking of candidate IRs in descending order of predicted performance, thereby directing the search process toward configurations that are more likely to yield optimal or near-optimal outcomes.
\tlr{Arafat ended using ChatGPT o1 here}

% our NCS model is designed to relatively rank a limited number of IRs generated by the supporting model, which must constrain subset of IRs to mostly fall within good regions of the search space.

% We simplify the learning task by focusing on distinguishing relative rather than absolute performance.
% To this end, we adopt a gradual pruning approach shown in Figure \ref{fig:ncs_overview}.

\subsubsection{Model Architecture}
% Our goal is not to predict one single class like traditional machine learning based predictions. Instead, our goal is to assign a score to each sample so that the best-performing samples get higher scores. To identify near-optimal configurations and assign them the right score, we employ multiple neural networks, each independently assigning an unweighted score to IRs proposed by GCTLA. We then sum up all the score provided by each model to generate the final score for a single sample. These scores enable us to rank IRs from high to low performance, guiding the search towards more promising candidates.

% To identify near-optimal configurations, we employ multiple convolutional neural networks, each independently assigning an unweighted score to IRs proposed by GCTLA.
% These scores enable us to rank IRs from high to low performance, guiding the search towards more promising candidates.
% By binning performance into a several classes, each with a separate neural network, we reduce the noise within each model's training dataset and incentivize predicting discrete classes during training and inference.

\tlr{This is the part to say we don't want the model to predict JUST ONE CLASS, but say it better.}

\tlr{Going back to challenges, we should explicitly motivate why we use the percentile approach -- ie the configurations that perform well should resemble one another a bit, but will evolve over time and may dip into low-performing region when conflicts occur.}

\tlr{Arafat began using ChatGPT o1 here}
The highest-performing code variants often share structural and semantic similarities in their intermediate representations.
%In the context of performance prediction for high-performance computing (HPC) code optimization, a central observation is that high-performing configurations often share structural or semantic similarities in their intermediate representations.
However, these similarities do not guarantee stable performance patterns, as performance may degrade due to conflicts between optimization choices and hardware-specific behaviors.
An unknown configuration may not be high-performing merely by merit of being similar to a known high-performing configuration.
\tlr{We could try to show this on Syr2k but probably not in time for SC25.}
This dynamic nature of the performance landscape---marked by local fluctuations and nonlinear interactions---necessitates a scoring mechanism that does not simply predict performance in isolation, but instead supports relative ranking across a spectrum of samples.
Our goal is to design a ranking-aware system capable of discerning nuanced performance potential across a diverse set of generated configurations.
By framing the learning task around the gradual separation of samples into performance tiers based on empirical speedup distributions, we allow the model to encode meaningful performance distinctions at multiple levels of granularity.
This approach also supports a more robust and flexible pruning strategy.
Instead of rigid classification, we enable a soft selection mechanism that favors configurations that resemble high-performance regions while tolerating some lapses in modeling fidelity.

To implement this ranking-based approach, we require a model architecture capable of capturing hierarchical patterns embedded in LLVM IR representations, while remaining efficient under data-scarce conditions.
To this end, we adopt a CNN architecture, which have demonstrated strong performance in settings with limited training data ~\cite{brigato2021close}.
Compared to other neural models, their small training and inference footprints
% time requirements for training and inference 
make them uniquely suitable for the minimal available data and tight latency constraints present in autotuning.

\tlr{We've already discussed:
\begin{itemize}
    \item Small train / inference time
    \item Small data requirement for training
\end{itemize}

We haven't discussed:
\begin{itemize}
    \item Ease of ``targeting'' to altered portion of IR (expected differentiable zone)
\end{itemize}
}

% However, the choice of CNN also introduces certain limitations.
Nonetheless, in our setting, the CNN architecture strikes a favorable balance: it is capable of learning meaningful structural features from compact datasets, while enabling efficient inference across large volumes of generated samples.
This makes it a practical and effective choice for our performance prediction pipeline.
\tlr{Arafat ended using ChatGPT o1 here}

To train this architecture effectively, we construct a dataset that efficiently captures the mapping between LLVM IR representations and their empirical performance. 
We collect configuration-level IR samples and their corresponding execution times from Ytopt\cite{wu2023ytopt}, using the same configurations that serve as the foundation for the supporting model's source-task training (details in Section \ref{sec:training_data_description}).
To focus on our analysis and reduce noise, we extract only the functions that are directly affected by configuration changes. 
Then we pre-process the IR contents by replacing the meta-data and other repetitive  information with specific keywords. This is similar to the approach in~\cite{barchi2019code}, but with a subtle refinement: we retain loop structures and array access patterns, given their critical role in influencing execution behavior and runtime variability. 
Rather than processing the raw IR text directly, we design a custom vocabulary to tokenize the LLVM IR into meaningful units. These tokenized sequences are then embedded into dense vector representations through an embedding layer, enabling the CNN to process structural and semantic relationships more effectively. Finally, the output of each CNN model is passed through a fully connected layer that maps the extracted features to a predicted score, enabling the ensemble to rank samples across continuous perforamnce tiers.

%\tlr{We may need a proper CNN model diagram to display this level of detail... omit and put in camera-ready/rebuttal?}
%These embeddings are then processed by a one-dimensional convolutional layer that applies multiple filters to extract local features.
%The convolutional outputs undergo a global max-pooling operation, which captures the most significant features across the sequence.
% The final, fully connected layer of each CNN maps learned features to its output class, allowing the ensemble to make predictions across the span of performance.

\subsubsection{Data Partitioning} \label{subsec:partitioning}

First, we calculate the speedup of each IR sample relative to the default configuration's execution time. 
Then, we partition the same empirical training dataset used to fit the GCTLA model for each benchmark into five performance tiers based on percentiles (50+, 60+, 70+, 80+ and 90+) of speedup.
Each of these five percentile groups represents a distinct performance level, allowing our models to learn meaningful distinctions between high- and low-performing configurations.
We label each sample positive if it belongs to a particular percentile group and negative otherwise.
In order to do that, we came up with the novel idea of ranking the input IRs based on the configuration that might provide the best performance.

\subsubsection{Training and Scoring}

As mentioned in subsection~\ref{subsec:partitioning}, the dataset is partitioned into five percentile-based groups according to speedup values.
A separate CNN model is trained for each group, ensuring specialization in different performance ranges.
During inference, each test sample is evaluated by all five models independently.
The final score for an individual sample is computed based on the sum of each of the five model predictions.
%If the $i^{th}$ model predicts positive for a configuration sample, the corresponding sample's score gets increased by one and zero otherwise.
This way, a particular IR can achieve a maximum score of five and a minimum score of zero. 
% as shown in Equation~\ref{eq:rank}, \tlr{Can we simplify this to say it's just the unweighted sum of all model classifications? The equation seems like overkill here.}

% \begin{equation}\label{eq:rank}
%     S=\sum_{i=1}^{5} M_i(x)
% \end{equation}

% where $M_i(x)$ represents the binary classification results for the $i^{th}$ model.

% This multi-model approach enables a pruning mechanism that significantly enhances the efficiency of the autotuning process.
% The rationale behind this design is to iteratively refine the search space by progressively filtering out low-performing configurations while retaining promising candidates for further evaluation.
The 50th percentile model acts as an initial filter that eliminates configurations that are likely to perform poorly.
As GCTLA is not supposed to sample many configurations that perform worse than this, we do not need lower classification levels and provide this as a safeguard against poor samples permitted by over-sampling.
The intermediate models (60th, 70th, and 80th percentiles) refine the selection by iteratively identifying configurations that exhibit increasing levels of performance, guiding the autotuner towards more promising regions.
Finally, the 90th percentile model pinpoints the best configurations, ensuring that only the most optimal candidates are retained for empirical evaluation.
\tlr{Expand upon this if needed $\rightarrow$}
Using multiple models in this manner is crucial because a single CNN model would struggle to effectively balance search space reduction with performance retention.
Since the search space is too large, if we relied on a single model or threshold, we risk either eliminating valuable configurations too early or retaining too many suboptimal ones, leading to inefficient tuning. 
% Why are we using the word 'prompt' here?

% This is repeating what you've already discussed -- is there supposed to be something different? I'm not convinced that this provides robustness over alternatives or that it is more appropriate for this task where we are data-constrained

% \subsection{Dataset Collection}
% For the embedding model, we have used C/C++ code samples from \arafat{name of the dataset}

% We use Bayesian optimization (BO) as a sample-efficient means to collect limited evaluations across the search space without prior data.

\section{Performance Evaluation} \label{sec:experiments} \subsection{Experimental Platform}

Our empirical evaluations are performed on a single Linux machine running Ubuntu 22.04.4 LTS on kernel version 5.15.0-78-generic.
The machine has 1TB DDR4 RAM with 2x AMD EPYC 7742 64-Core Processor (128 logical cores) and 8x NVIDIA A100 GPUS for a total of 320GB GPU memory.

Several of the benchmarks we tune utilize Polly~\cite{polly}, a polyhedral loop optimizer for LLVM, to expose source-code-level optimizations for our experiments.
The Polly extensions used in this work are publicly available via a customized compiler based on Clang version 13.0.0, provided by the SOLLVE LLVM project~\cite{SOLLVE}.

\tlr{This machine is less powerful than the evaluation machine, I think we can omit describing it as we'd probably just reduce the NCS training time a pinch over here:

CNN model training was performed on a single Linux machine running SUSE Linux Enterprise Server (SLES) 15 SP5.
The machine has 512GB DDR4 RAM with AMD Zen 3 (Milan) 32-Core Processor and 4x NVIDIA A100-SXM4-40GB GPUs for a total of 160GB GPU memory. \arafat{The choice of this machine for training was driven by its specialized configuration, which is well-suited for the computational demands of deep learning workloads and to maintain an efficient workflow to prevent interference between deep learning model training and autotuning experiments.}
}

\subsection{Tunable Applications}

We focus our tuning on three key kernels from the Polybench/C~\cite{polybench} collection of benchmarks and three additional applications selected from the Exascale Computing Project proxy applications.
Table~\ref{tbl:GC_spaces} provides a general overview of the size of each benchmark's search space as well as the number of unique configurations the GCTLA generates versus its expected budget for few-shot tuning as reported in Randall et al~\cite{GC_TLA}.
The few-shot budget of GCTLA is minimal compared to the search space size, but the total number of unique configurations is too large to empirically evaluate.
We explicitly move beyond the extremely limited number of evaluations that GCTLA is designed for, but not so far as to exhaust the full extent of the model's variability.

\begin{table}[t]
	\begin{center}
		\caption{Search Space Sizes and GCTLA Coverage} \label{tbl:GC_spaces}
        {\scriptsize
		\begin{tabular}{|l|c|l|l|l|}
			\hline
			\multirow{3}*{Benchmark} & \multirow{3}*{\# Params}& \multirow{3}*{\# Configurations} & GCTLA & GCTLA \# \\
             & & & Few-Shot & Unique \\
             & & & Budget & Configurations \\
			\hline
			3mm & 10 & 376,320 & 30 & $\approx$ 2,500 \\
			%Covariance & 5 & 5,324 & 30 & $\approx$ 110 \\
			%Floyd--Warshall & 5 & 5,324 & 15 & $\approx$ 1,800 \\
			Heat3d & 6 & 10,648 & 8 & $\approx$ 1,600 \\
			%LU & 5 & 5,324 & 30 & $\approx$ 210 \\
			Syr2k & 6 & 10,648 & 3 & $\approx$ 800 \\\hline
			AMG & 9 & 1,180,980 & 5 & $\approx$ 108,500 \\
			RSBench & 9 & 5,196,312 & 3 & $\approx$ 316,800 \\
			%XSBench & 8 & 577,368 & 7 & $\approx$ 77,500 \\
			SW4Lite & 8 & 4,752 & 15 & $\approx$ 1,800 \\
			\hline
		\end{tabular}
        }
	\end{center}
\end{table}

\subsubsection{Polybench/C Kernels}

In prior work, the Polybench/C kernels demonstrated some promise for the GCTLA as a generative autotuners, however its explicit few-shot design prevents it from continuing searches to discover better optima within the search spaces.
The three Polybench/C kernels we focus on are 3MM, Heat3d and Syr2k; the tuning parameters for these search spaces are listed in Table~\ref{tbl:polybench_params}.

The 3MM kernel previously demonstrated excellent results for the GCTLA on its own in short-term searches, representing an opportunity for our technique to extend the search in a productive manner.
The Heat3d kernel previously performed the worst for the GCTLA relative to other tuning techniques, representing an opportunity for our technique to bolster the generative effectiveness when discrimination is difficult.
Finally, the Syr2k kernel performed well but also has exhaustive data on two input data scales, representing a richer evaluation subject that can provide deeper insights to the effectiveness of our combined technique compared to its component parts or other tuners.

\begin{table*}[h]
    \begin{center}
        \caption{Parameters used to tune Polybench Kernels. Values within brackets indicate the options available for an independent parameter, and a list of brackets represents multiple independent parameters.}
        \label{tbl:polybench_params}
        {\scriptsize
        \begin{tabular}{|l|c|c|c|c|c|c|}
        \hline
        Parameter & 3MM & Heat3d & Syr2k \\\hline
        \multirow{2}{*}{Tile Sizes} & [4-2048], [4-2048], & [4-128], [4-2048], & [4-128], [4-2048], \\
         & [4-2048] & [4-256] & [4-256] \\\hline
        Loop Interchange & [Yes, N/A] & [Yes, N/A] & [Yes, N/A] \\\hline
        Array Packing & [Yes, N/A] $\times$ 6 & [Yes, N/A] $\times$ 2 & [Yes, N/A] $\times$ 2 \\\hline
        \end{tabular}
        }
    \end{center}
\end{table*}

\subsubsection{ECP Proxy Applications}

The Polybench/C benchmarks are not trivial to tune but generally yield greater speedups than sophisticated applications that are already somewhat optimized.
As such, we follow prior work and also consider our tuner's effectiveness on the AMG, RSBench and SW4Lite ECP Proxy Applications.
The tuning parameters for these search spaces are listed in Table~\ref{tbl:exp_params}.

% * ECP AMG - GC's best prior result on ECP
% * ECP RSBench - GC arguably worst result for ECP
% * ECP SW4Lite - Architecture independence

Similar to 3MM from the Polybench/C collection, the AMG application yielded the best results for the GCTLA on its own, representing opportunity to extend searches to yield greater performance improvements.
The RSBench and XSBench applications are highly similar, so in the interest of time we focus on RSBench, which was previously shown to be slightly more challenging to speed up via tuning.
Finally we include the GPU-enabled SW4Lite application to demonstrate tuning performance on different compute architectures.

\begin{table*}[h]
    \begin{center}
        \caption{Parameters used to tune ECP proxy applications. %Brackets represent options for independent parameters.
        }
        \label{tbl:exp_params}
        {\scriptsize
        \begin{tabular}{|l|c|c|c|c|}
        \hline
        Parameter & AMG & RSBench & SW4Lite \\\hline
        \multirow{2}{*}{Tile Sizes} & [10-200], [2-256], & \multirow{2}{*}{[2-256], [2-256]} & \multirow{2}{*}{--} \\
         & [2-256], [10-200] & & \\\hline
        \multirow{2}{*}{Optional Parameters} & \multirow{2}{*}{Parallel For} & \multirow{2}{*}{Parallel For} & Parallel For, Nowait, \\
         & & & MPI\_Barrier \\\hline
        Parallel For Schedule & -- & [100-2000], [10-200] & [dynamic, static] \\\hline
        \multirow{2}{*}{Unrolling Options} & \multirow{2}{*}{[unroll, N/A]} & \multirow{2}{*}{[unroll, N/A]} & [unroll (6) \footnotemark[2], unroll, \\
         & & & no-unroll] \\\hline
        \# Threads & [4-8] & [2-256] & [2-256] \\\hline
        \multirow{3}{*}{KMP Affinity} & [compact, scatter, & [compact, scatter, & \multirow{3}{*}{--} \\
         & balanced] & balanced, none, & \\
         & & disabled, explicit] & \\\hline
        OMP Proc Bind & -- & -- & [close, spread, master] \\\hline
        OMP Places & [core, threads, sockets] & [core, threads, sockets] & [core, threads, sockets] \\\hline
        \end{tabular}
        }
    \end{center}
\end{table*}

\subsection{Initial Dataset and Problem Formulation} \label{sec:training_data_description}

As our work focuses on extended search capabilities, we preserve the same initial dataset as previously used in Randall et al~\cite{GC_TLA}.
These datasets are composed of three different input data sizes (small, medium, and large) for each kernel or application, with 200 empirical evaluations per data size (a total of 600 empirical observations per benchmark).
The observations are selected by YTOPT~\cite{wu2023ytopt}, however we additionally collect the IR of each configured benchmark for use with NCS.

For fairness, we present the exact same training data in its most appropriate form to compared techniques that can benefit from prior data from related tasks.
Each search begins without any data relating to the target task, representing a fresh start to optimize the related performance objective.
All evaluations are performed on two new tasks, an interpolated size (Small-Medium, or SM, between small and medium source tasks) and an extrapolated size (Extra-Large, or XL, larger than the large source task) on all benchmarks.
To minimize variance in system performance and randomness of tuning techniques, we report all empirical evaluations as the average of three executions and repeat all tuning runs with three separate randomized seeds.

For each new task, we permit a maximum of 300 empirical evaluations or 24 hours to be spent tuning, whichever comes first.

\subsection{Training and Inference}\label{subsec:train_and_inf}

Training the GCTLA model requires no more than a few seconds, and sampling time is trivial.
We trained all five NCS models with configurations generated by YTOPT for small, medium, and large input sizes.
Due to the small size of the networks and limited training data, training the NCS ensemble in this way takes about four minutes for each benchmark; this forms the majority of the overhead of our technique.
We have used the Adam Optimizer for training, and each model was trained for 30 epochs.
Similar to GCTLA, NCS's inference time for all sampled IRs is trivial.
Including to the overhead of compiling, the training+inference time of our technique ``costs'' no more than ten evaluations on any benchmark.
\tlr{$\leftarrow$ Should we explicitly equate this to an empirical observation tax per benchmark? Could be an extra column in Table 1 (search space sizes and GCTLA Coverage)}

\subsection{Compared Techniques}

We compare our techniques to the GCTLA without NCS in addition to state-of-the-art autotuners BLISS~\cite{bliss}, OpenTuner~\cite{OpenTuner}, and GPTune~\cite{GPTuneBand,GPTuneCrowd}.

BLISS is a meta-technique that smoothly interpolates between lightweight surrogate models to effectively identify the best learning strategy during the autotuning process.

OpenTuner is an autotuning framework that supports over twenty-five separate search algorithms and can be extended to support additional search algorithms.
Due to the excessive cost of demonstrating the performance of all of OpenTuner's techniques, we utilize the exhaustive tuning data for the Syr2k SM and XL problems to determine the best-performing technique.
Empirically, we find the AUCBanditMetaTechniqueB, ga-PX and NormalGreedyMutation10 techniques to be highly competitive with one another in an initial exploration of available algorithms, and elect to utilize the NormalGreedyMutation10 search technique for the remaining benchmarks.
Henceforth, the NormalGreedyMutation10 algorithm executed by OpenTuner is referred to as ``OpenTuner'' for brevity.
% Show some results here or in an appendix for why we pick NormalGreedyMutation10

GPTune is a machine-learning multi-task transfer autotuning technique built upon Gaussian Processes.
As a transfer-enabled autotuning technique, GPTune is capable of reusing the same training information utilized in our own searches and is accordingly set up to utilize this same data.
The authors of GPTune also contributed the GPTuneCrowd~\cite{GPTuneCrowd} database strategy to facilitate knowledge sharing and empower the community with collective autotuning efforts.
% GPTune struggles to fit very large evaluation traces, which it has to do for each predictive evaluation after its randomly selection burn-in period ends (per the authors' advice: half of all permitted evaluations).

In order to clearly demonstrate the capabilities of NCS, we perform all searches without NCS and reorder the searches afterwards.
In addition to clearly displaying the advantages and shortcomings of NCS, this allows us to apply NCS to iterative techniques that cannot generate many samples at once.
This analysis allows us to determine if other generative samplers than GCTLA can be expected to perform well with NCS, however we are not aware of any highly suitable alternatives at the time of writing.

\subsection{Experimental Results}

\ali{It would be helpful to show the differences in the number of trials needed to achieve the best speedup, with and without NCS, in a small additional diagram to emphasize the differences. -- \tlr{candlestick/whisker plot maybe? -- something that is very simple (average difference vs GC+NCS), maybe just pick a few to highlight.}}

\begin{table}[h]
\caption{Speedups of Polybench/C Apps}
\label{tbl:polybench_speedup}
\centering
\scriptsize
\begin{tabular}{|c|c|c|c|c|}
\hline
\multirow{3}*{Benchmark} & Target & \multirow{3}*{Search Tool} & (First 100, Best) & Average \\
 & Size & & Speedup & \# Trials \\
 & & & & to Best \\\hline

  &   & Default & (1.000, 1.000) & 1 \\
  &   & GCTLA & (\textbf{3.645}, \textbf{3.721}) & 201.67 \\
  &   & GPTune & (3.042, 3.042) & 77 \\
  &   & BLISS & (3.101, 3.143) & 59.33 \\
SyR2K & SM & OpenTuner & (3.582, 3.582) & 69.33 \\\cline{3-5}
  &   & GC+NCS (ours) & (\textbf{3.645}, 3.645) & 16 \\\cline{3-5}
  &   & GPTune+NCS & (3.027, 3.042) & 39.33 \\
  &   & BLISS+NCS & (2.994, 3.143) & 16.33 \\
  &   & OpenTuner+NCS & (3.374, 3.582) & 36.33 \\
\rowcolor{gray!25}  &   & Default & (1.000, 1.000) & 1 \\
\rowcolor{gray!25}  &   & GCTLA & (1.177, \textbf{1.189}) & 245.33 \\
\rowcolor{gray!25}  &   & GPTune & (1.037, 1.037) & 57.33 \\
\rowcolor{gray!25}  &   & BLISS & (1.163, 1.163) & 46.67 \\
\rowcolor{gray!25} SyR2K & XL & OpenTuner & (\textbf{1.178}, 1.185) & 114.67 \\\cline{3-5}
\rowcolor{gray!25}  &   & GC+NCS (ours) & (1.175, \textbf{1.189}) & 18 \\\cline{3-5}
\rowcolor{gray!25}  &   & GPTune+NCS & (1.032, 1.037) & 32.67 \\
\rowcolor{gray!25}  &   & BLISS+NCS & (1.163, 1.163) & 25 \\
\rowcolor{gray!25}  &   & OpenTuner+NCS & (\textbf{1.179}, 1.185) & 65.33 \\
\hline
  &   & Default & (1.000, 1.000) & 1 \\
  &   & GCTLA & (5.458, \textbf{5.894}) & 150 \\
  &   & GPTune & (\textbf{5.857}, 5.881) & 76 \\
  &   & BLISS & (4.327, 4.566) & 72.67 \\
3MM & SM & OpenTuner & (4.664, 5.010) & 85.67 \\\cline{3-5}
  &   & GC+NCS (ours) & (5.445, \textbf{5.894}) & 95 \\\cline{3-5}
  &   & GPTune+NCS & (5.424, 5.881) & 7.33 \\
  &   & BLISS+NCS & (4.503, 4.566) & 44.67 \\
  &   & OpenTuner+NCS & (5.010, 5.010) & 36.67 \\
\rowcolor{gray!25}  &   & Default & (1.000, 1.000) & 1 \\
\rowcolor{gray!25}  &   & GCTLA & (38.196, \textbf{38.726}) & 183 \\
\rowcolor{gray!25}  &   & GPTune & (23.856, 23.856) & 35.33 \\
\rowcolor{gray!25}  &   & BLISS & (17.841, 32.314) & 80 \\
\rowcolor{gray!25}3MM & XL & OpenTuner & (37.907, 38.348) & 105.33 \\\cline{3-5}
\rowcolor{gray!25}  &   & GC+NCS (ours) & (38.315, \textbf{38.726}) & 106.67 \\\cline{3-5}
\rowcolor{gray!25}  &   & GPTune+NCS & (23.459, 23.856) & 7 \\
\rowcolor{gray!25}  &   & BLISS+NCS & (31.194, 32.314) & 114 \\
\rowcolor{gray!25}  &   & OpenTuner+NCS & (\textbf{38.348}, 38.348) & 66.67 \\
\hline
  &   & Default & (1.000, 1.000) & 1 \\
  &   & GCTLA & (2.077, \textbf{2.357}) & 216.33 \\
  &   & GPTune & (2.215, 2.215) & 66 \\
  &   & BLISS & (2.233, 2.233) & 38.33 \\
Heat3D & SM & OpenTuner & (2.091, 2.091) & 33.67 \\\cline{3-5}
  &   & GC+NCS (ours) & (\textbf{2.357}, \textbf{2.357}) & 33 \\\cline{3-5}
  &   & GPTune+NCS & (2.215, 2.215) & 11.33 \\
  &   & BLISS+NCS & (2.233, 2.233) & 27 \\
  &   & OpenTuner+NCS & (2.091, 2.091) & 40.33 \\
\rowcolor{gray!25}  &   & Default & (1.000, 1.000) & 1 \\
\rowcolor{gray!25}  &   & GCTLA & (\textbf{3.341}, \textbf{3.344}) & 123.67 \\
\rowcolor{gray!25}  &   & GPTune & (3.181, 3.181) & 59.67 \\
\rowcolor{gray!25}  &   & BLISS & (3.329, 3.338) & 168 \\
\rowcolor{gray!25}Heat3D & XL & OpenTuner & (3.314, 3.314) & 88 \\\cline{3-5}
\rowcolor{gray!25}  &   & GC+NCS (ours) & (3.339, \textbf{3.344}) & 149.67 \\\cline{3-5}
\rowcolor{gray!25}  &   & GPTune+NCS & (3.181, 3.181) & 6 \\
\rowcolor{gray!25}  &   & BLISS+NCS & (3.336, 3.338) & 37 \\
\rowcolor{gray!25}  &   & OpenTuner+NCS & (3.314, 3.314) & 5 \\
\hline

\hline
\end{tabular}
\end{table}

\begin{table}[h]
\caption{Speedups of ECP Apps}
\label{tbl:ecp_speedup}
\centering
\scriptsize
\begin{tabular}{|c|c|c|c|c|}
\hline
\multirow{3}*{Benchmark} & Target & \multirow{3}*{Search Tool} & (First 100, Best) & Average \\
 & Size & & Speedup & \# Trials \\
 & & & & to Best \\\hline

  &   & Default & (1.000, 1.000) & 1 \\
  &   & GCTLA & (0.668, 0.683) & 145 \\
  &   & GPTune & (0.978, 0.978) & 84 \\
  &   & BLISS & (0.602, 0.608) & 123.67 \\
AMG & SM & OpenTuner & (\textbf{1.509}, \textbf{1.509}) & 29.67 \\\cline{3-5}
  &   & GC+NCS (ours) & (0.679, 0.683) & 80.67 \\\cline{3-5}
  &   & GPTune+NCS & (0.978, 0.978) & 54 \\
  &   & BLISS+NCS & (0.607, 0.608) & 69 \\
  &   & OpenTuner+NCS & (1.507, \textbf{1.509}) & 89.67 \\
\rowcolor{gray!25}  &   & Default & (1.000, 1.000) & 1 \\
\rowcolor{gray!25}  &   & GCTLA & (0.928, 0.930) & 226.67 \\
\rowcolor{gray!25}  &   & GPTune & (0.887, 0.887) & 24.33 \\
\rowcolor{gray!25}  &   & BLISS & (0.893, 0.914) & 186.33 \\
\rowcolor{gray!25}AMG & XL & OpenTuner & (\textbf{1.046}, \textbf{1.046}) & 41.33 \\\cline{3-5}
\rowcolor{gray!25}  &   & GC+NCS (ours) & (0.927, 0.930) & 155.33 \\\cline{3-5}
\rowcolor{gray!25}  &   & GPTune+NCS & (0.887, 0.887) & 27.67 \\
\rowcolor{gray!25}  &   & BLISS+NCS & (0.912, 0.914) & 85.67 \\
\rowcolor{gray!25}  &   & OpenTuner+NCS & (\textbf{1.046}, \textbf{1.046}) & 52.33 \\
\hline
  &   & Default & (1.000, 1.000) & 1 \\
  &   & GCTLA & (0.749, 0.749) & 51.33 \\
  &   & GPTune & (1.998, \textbf{3.290}) & 81.33 \\
  &   & BLISS & (0.916, 1.079) & 146 \\
RSBench & SM & OpenTuner & (1.100, 1.132) & 131.33 \\\cline{3-5}
  &   & GC+NCS (ours) & (0.729, 0.749) & 162.33 \\\cline{3-5}
  &   & GPTune+NCS & (\textbf{3.290}, \textbf{3.290}) & 40.67 \\
  &   & BLISS+NCS & (0.894, 1.079) & 140.33 \\
  &   & OpenTuner+NCS & (1.132, 1.132) & 87.67 \\
\rowcolor{gray!25}  &   & Default & (1.000, 1.000) & 1 \\
\rowcolor{gray!25}  &   & GCTLA & (0.528, 0.529) & 129.67 \\
\rowcolor{gray!25}  &   & GPTune & (0.765, 0.765) & 71.67 \\
\rowcolor{gray!25}  &   & BLISS & (1.002, \textbf{1.052}) & 91 \\
\rowcolor{gray!25}RSBench & XL & OpenTuner & (0.826, 0.826) & 72 \\\cline{3-5}
\rowcolor{gray!25}  &   & GC+NCS (ours) & (0.525, 0.529) & 165 \\\cline{3-5}
\rowcolor{gray!25}  &   & GPTune+NCS & (0.759, 0.765) & 52 \\
\rowcolor{gray!25}  &   & BLISS+NCS & (\textbf{1.052}, \textbf{1.052}) & 54.67 \\
\rowcolor{gray!25}  &   & OpenTuner+NCS & (0.826, 0.826) & 18.67 \\
\hline
  &   & Default & (\textbf{1.000}, \textbf{1.000}) & 1 \\
  &   & GCTLA & (0.956, 0.964) & 210 \\
  &   & GPTune & (0.929, 0.929) & 46.67 \\
  &   & BLISS & (0.778, 0.778) & 31 \\
SW4Lite & SM & OpenTuner & (0.945, 0.945) & 73 \\\cline{3-5}
  &   & GC+NCS (ours) & (0.956, 0.964) & 28.33 \\\cline{3-5}
  &   & GPTune+NCS & (0.929, 0.929) & 64.67 \\
  &   & BLISS+NCS & (0.768, 0.778) & 137 \\
  &   & OpenTuner+NCS & (0.945, 0.945) & 23 \\
\rowcolor{gray!25}  &   & Default & (1.000, 1.000) & 1 \\
\rowcolor{gray!25}  &   & GCTLA & (\textbf{1.039}, \textbf{1.040}) & 194.33 \\
\rowcolor{gray!25}  &   & GPTune & (1.034, 1.034) & 62.33 \\
\rowcolor{gray!25}  &   & BLISS & (0.943, 0.944) & 154.33 \\
\rowcolor{gray!25}SW4Lite & XL & OpenTuner & (1.032, 1.032) & 40.33 \\\cline{3-5}
\rowcolor{gray!25}  &   & GC+NCS (ours) & (1.039, \textbf{1.040}) & 64.33 \\\cline{3-5}
\rowcolor{gray!25}  &   & GPTune+NCS & (1.034, 1.034) & 47.67 \\
\rowcolor{gray!25}  &   & BLISS+NCS & (0.942, 0.944) & 167 \\
\rowcolor{gray!25}  &   & OpenTuner+NCS & (1.032, 1.032) & 53.67 \\
\hline

\hline
\end{tabular}
\end{table}

\begin{figure*}
\centering
\begin{subfigure}{0.45\textwidth}
    \includegraphics[width=\textwidth]{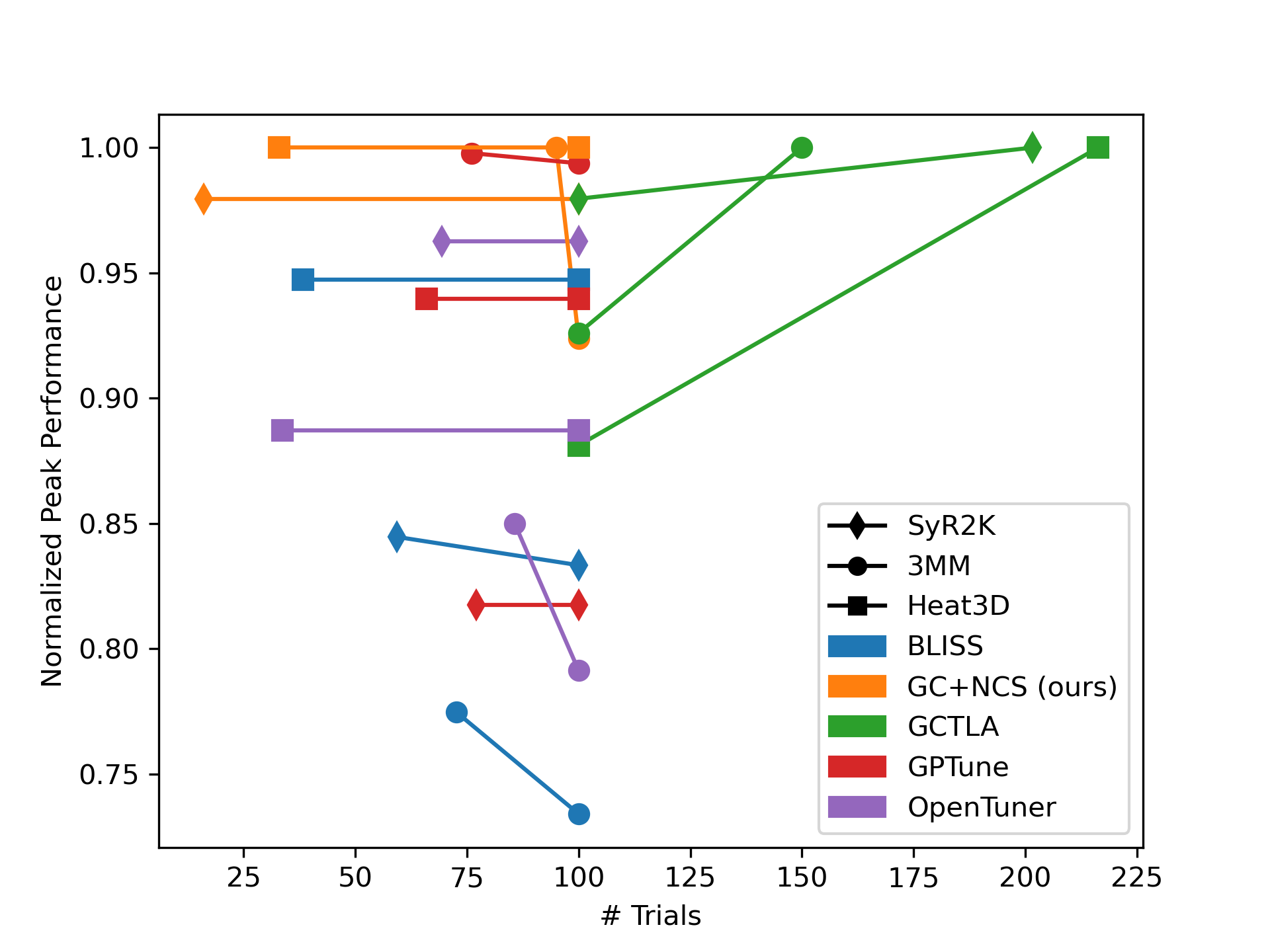}
    \caption{Polybench/C SM target tasks}
    \label{fig:general_SM_poly}
\end{subfigure}
\hfill
\begin{subfigure}{0.45\textwidth}
    \includegraphics[width=\textwidth]{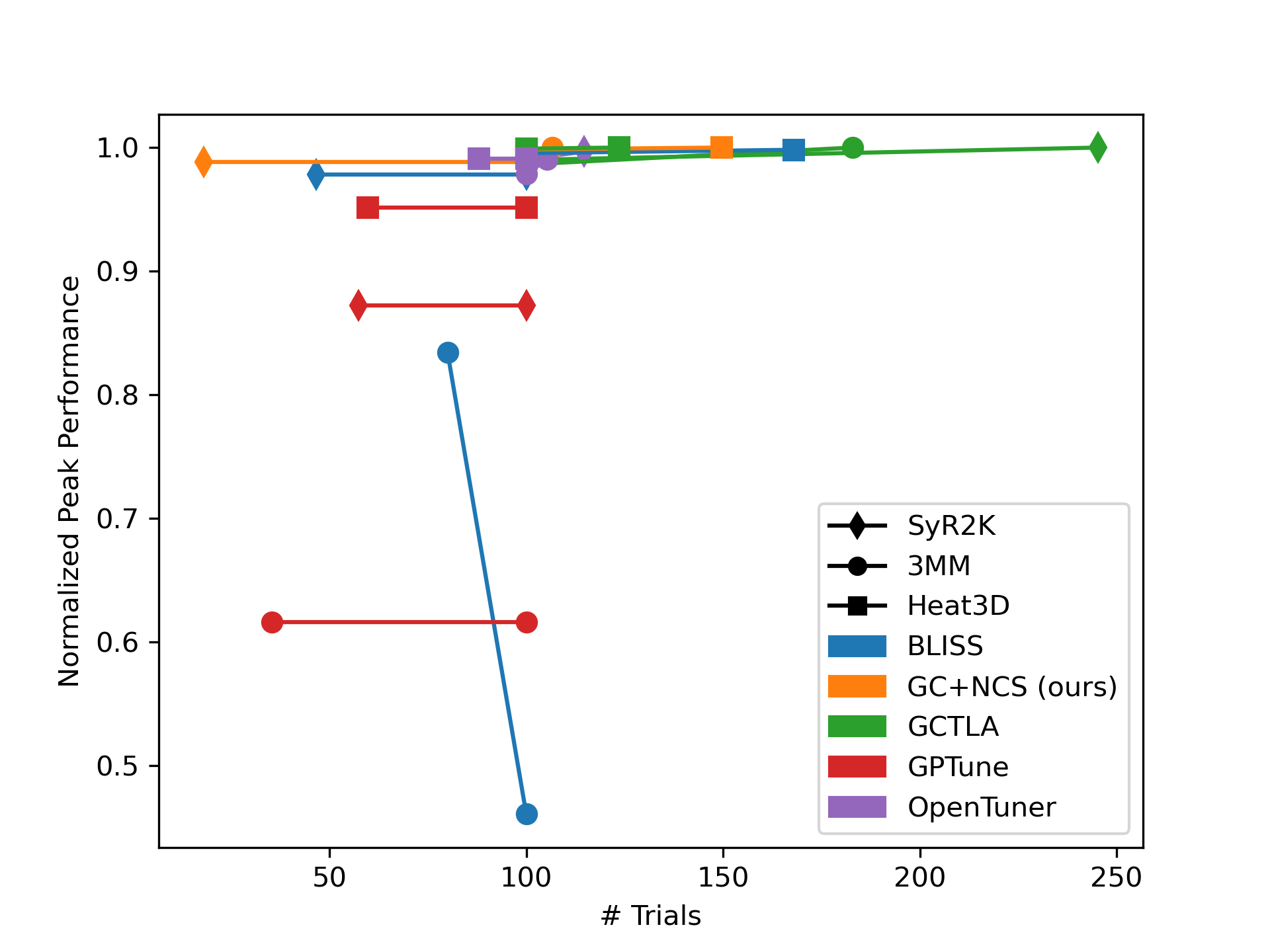}
    \caption{Polybench/C XL target tasks}
    \label{fig:general_XL_poly}
\end{subfigure}

\begin{subfigure}{0.45\textwidth}
    \includegraphics[width=\textwidth]{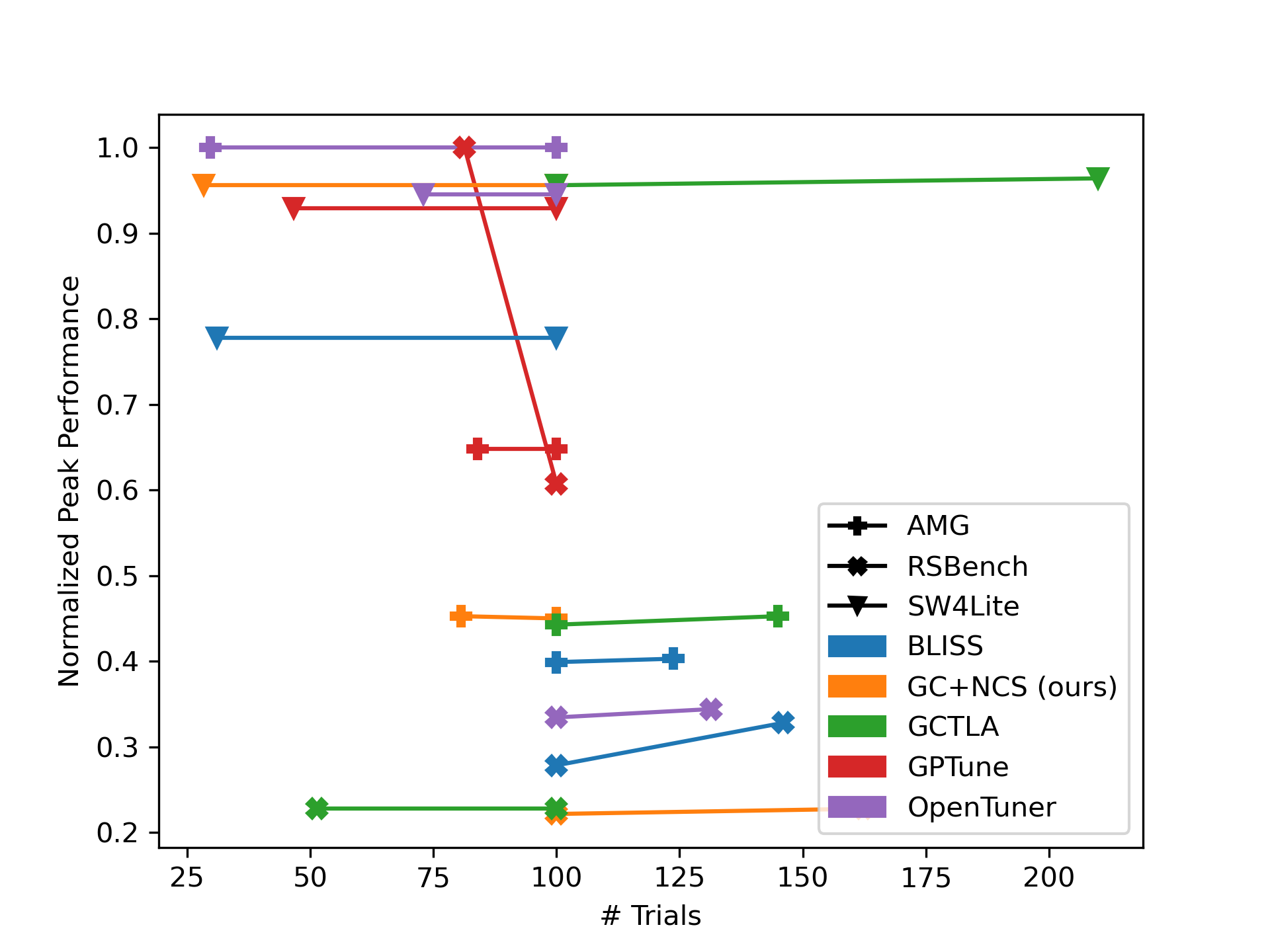}
    \caption{ECP SM target tasks}
    \label{fig:general_SM_ecp}
\end{subfigure}
\hfill
\begin{subfigure}{0.45\textwidth}
    \includegraphics[width=\textwidth]{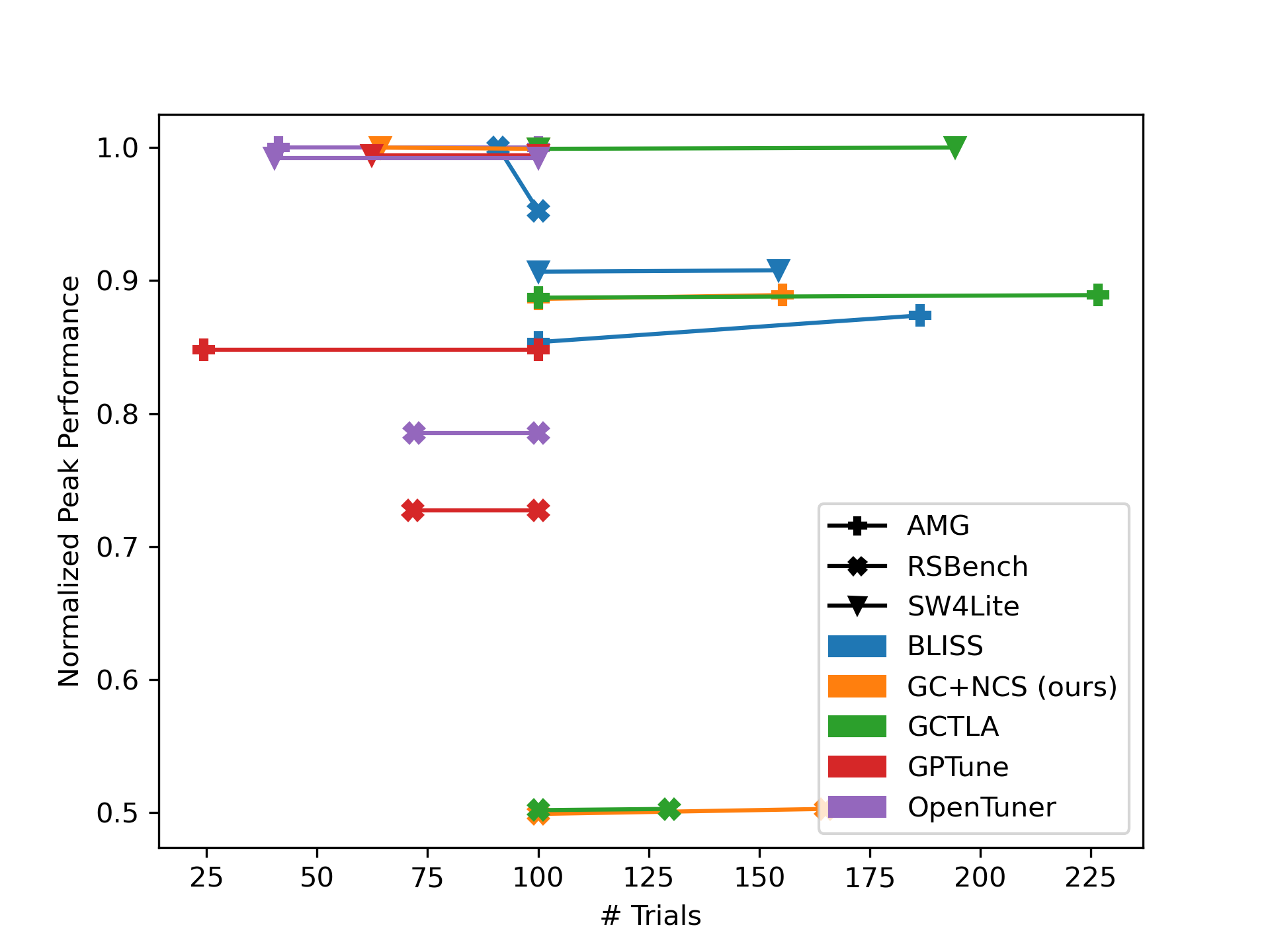}
    \caption{ECP XL target tasks}
    \label{fig:general_XL_ecp}
\end{subfigure}

\caption{Each line connects a search's average best result (based on its average trial number) to its average best result observed within 100 evaluations; observing the highest normalized performance within the fewest trials (up and to the left) is optimal.
\tlr{THIS FIGURE COULD BE BETTER!!}
}
\label{fig:general_results}
\end{figure*}

Tables~\ref{tbl:polybench_speedup} and~\ref{tbl:ecp_speedup} demonstrate the general results from our experiments.
We also provide a visual representation of the results in Figure~\ref{fig:general_results}.

\subsubsection{Immediate Efficacy}

We consider the first one hundred evaluations to be the ``immediate-term'' autotuning that is most consistent with prior work.

Across Polybench/C benchmarks in Table~\ref{tbl:polybench_speedup}, the GCTLA produces the best results of all search techniques.
With the help of NCS, we can frequently access the optimal result from GCTLA within the first 100 evaluations, reducing the average number of trials from 201.67 to 16 on the SyR2K SM benchmark, 61.67\% of the overall evaluation budget.
\tlr{ABSTRACT FIGURE}
Notably, within the first 100 trials, GC+NCS is reliably able to outperform most other autotuners' best results within the full search duration.
This demonstrates the capability of generative predictive autotuning to dramatically improve search performance.

\subsubsection{Ultimate Efficacy}

We also include the ultimate efficacy, or best result of each technique, to highlight the capabilities of our compared approaches within the full scope of its evaluations.
We were surprised to find that iterative techniques were generally unable to exceed 200 evaluations within the 24 hour allotment, which GCTLA and NCS were able to utilize their full 300 evaluations across all experiments.

We find that the ECP applications generally provide much larger dividends for iterative techniques, as their active learning capabilities are better suited to continue learning the complex performance relationships beyond the hints provided by source task training data.
We can see that NCS generally still works well with GCTLA, however some benchmarks such as RSBench prove challenging for NCS as well, resulting in the worst of both worlds.

Despite these setbacks, we can see that when NCS is applied after-the-fact to iterative autotuners, it still reduces the number of trials needed to discover the optimum, which supports our technique's applicability to other searches.
We expect that this implies NCS is well-suited for usage with other generative autotuners than GCTLA, and may be able to provide value on these more challenging benchmarks to generative techniques that can make use of iterative feedback. \tlr{May be a conclusion statement rather than experiments.}

\subsection{Exhaustive Searches: Syr2k}\label{sec:exhaustive}

\begin{figure*}
\centering
\begin{subfigure}{0.45\textwidth}
    \includegraphics[width=\textwidth]{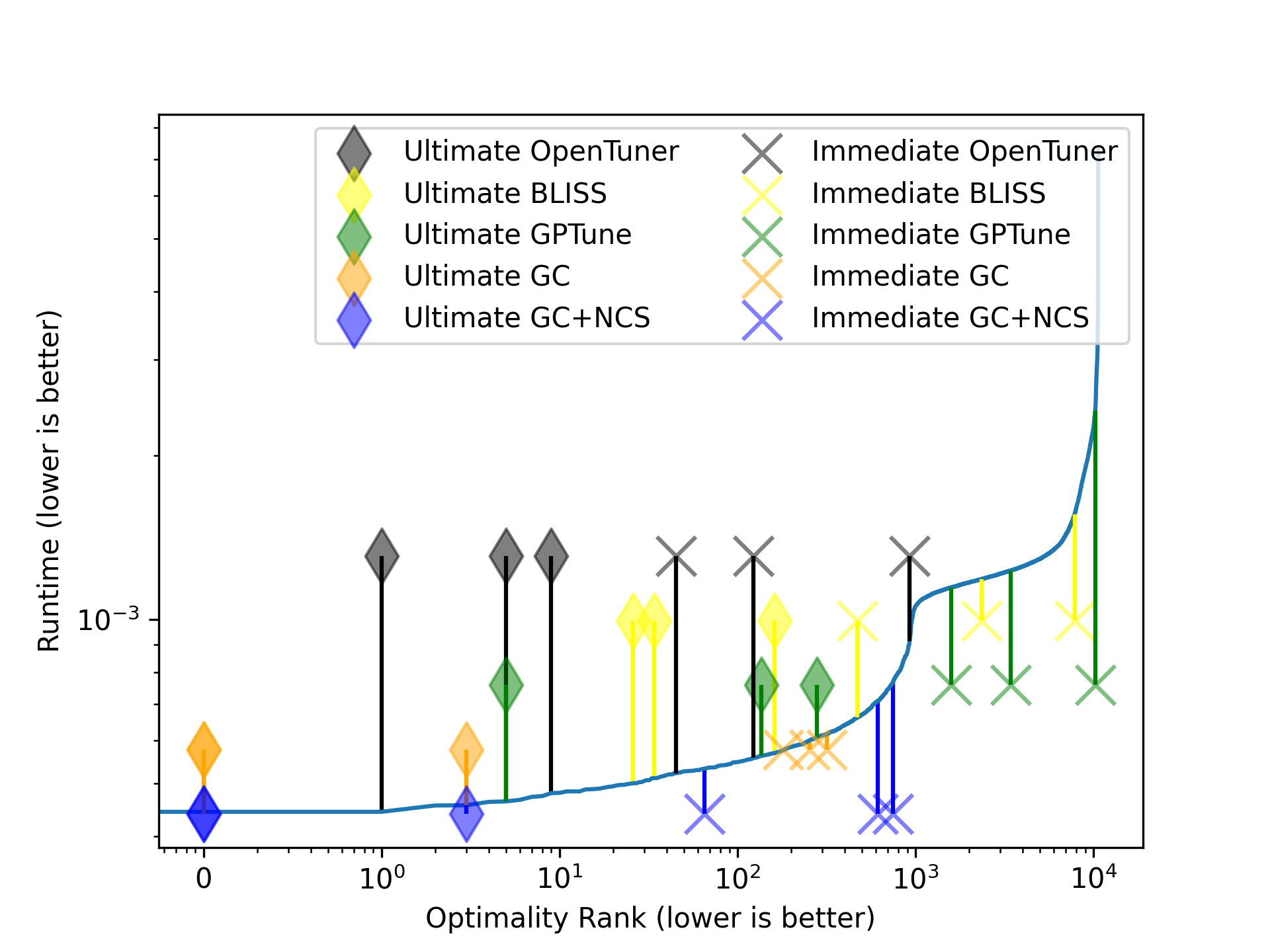}
    \caption{SM target task}
    \label{fig:syr2k_global_SM}
\end{subfigure}
\hfill
\begin{subfigure}{0.45\textwidth}
    \includegraphics[width=\textwidth]{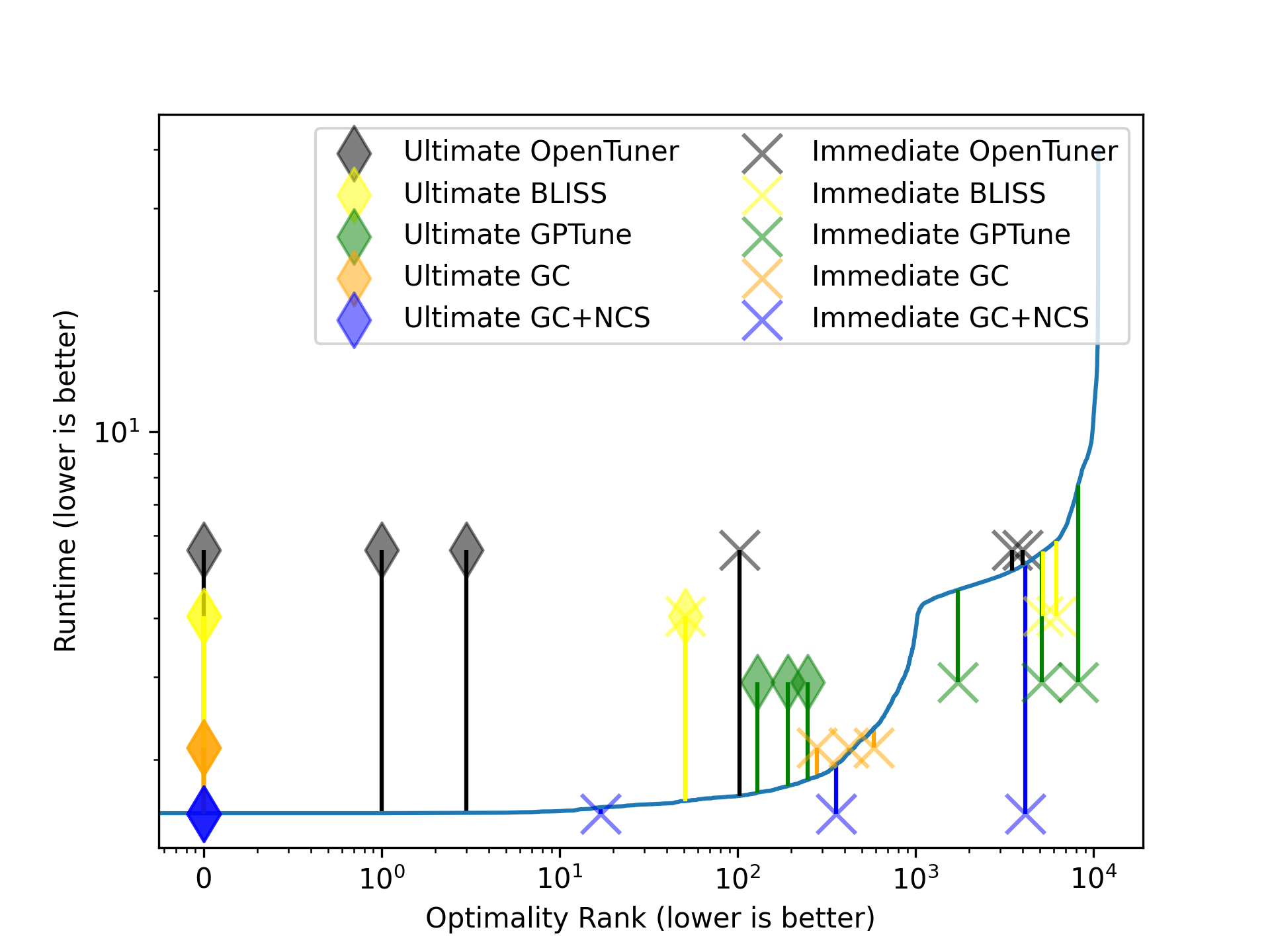}
    \caption{XL target task}
    \label{fig:syr2k_global_XL}
\end{subfigure}
\caption{SyR2K overall search optimality per-seed shown on a log-log scale.
Ideally, searches converge at the lowest rank.
%GPTune and BLISS often converge to comparatively poor best-case results, but GC+NCS prevents low-quality results from being searched.
}
\end{figure*}

We utilize exhaustive data for performance on all configurations of the Syr2k benchmark to demonstrate the global effectiveness of tuning and full-duration effectiveness of tuning techniques.
In Figures~\ref{fig:syr2k_global_SM} and~\ref{fig:syr2k_global_XL}, we see the global optimality achieved by different tuning techniques.
Notably, most techniques are not capable of performing complete searches, denoted by diamonds at their final evaluation.
For techniques utilizing GCTLA, this represents exhaustion of all unique samples that can be reasonably generated.
Because the model is unlikely to produce any of these excluded results, they are considered inaccessible to the search techniques.
Similar problems occur for OpenTuner, which automatically detects duplicate generations from its model and early-terminates searches that converge.
It is possible that similar problems occur for GPTune, however due to the repeated training costs for refitting all GPs at each step of the search, we find that this training cost begins to surpass reasonable limitations as the search length extends past 100 total evaluations.
This 100-evaluation limit falls within the previously published use cases for the framework, so while longer searches are theoretically possible, we find that the increasing overhead of refitting GPTune makes longer-term tuning efforts inadvisable.
We observe less difficulties in long-term searches with BLISS, however its surrogate evaluations are never re-examined, meaning that it may not discover the actual best performance when mispredictions occur, even if given additional budget to pursue extra evaluations.

\tlr{
This is not regular draft content, but I guess it could be in some form or another -- just not this one.
\begin{itemize}
    \item How difficult is the problem to solve? Pagerank centrality analysis
    \item How well do GC and MGA do independently? Including full time for E2E
    \item How much improvement do we recognize together?
    \item How well do OpenTuner/BLISS do independently? Can they be used with GC or with MGA?
Exploring limits of what we can achieve
\end{itemize}
}

\subsection{Details of NCS Effectiveness}

\begin{figure}
    \centering
    \includegraphics[width=\linewidth]{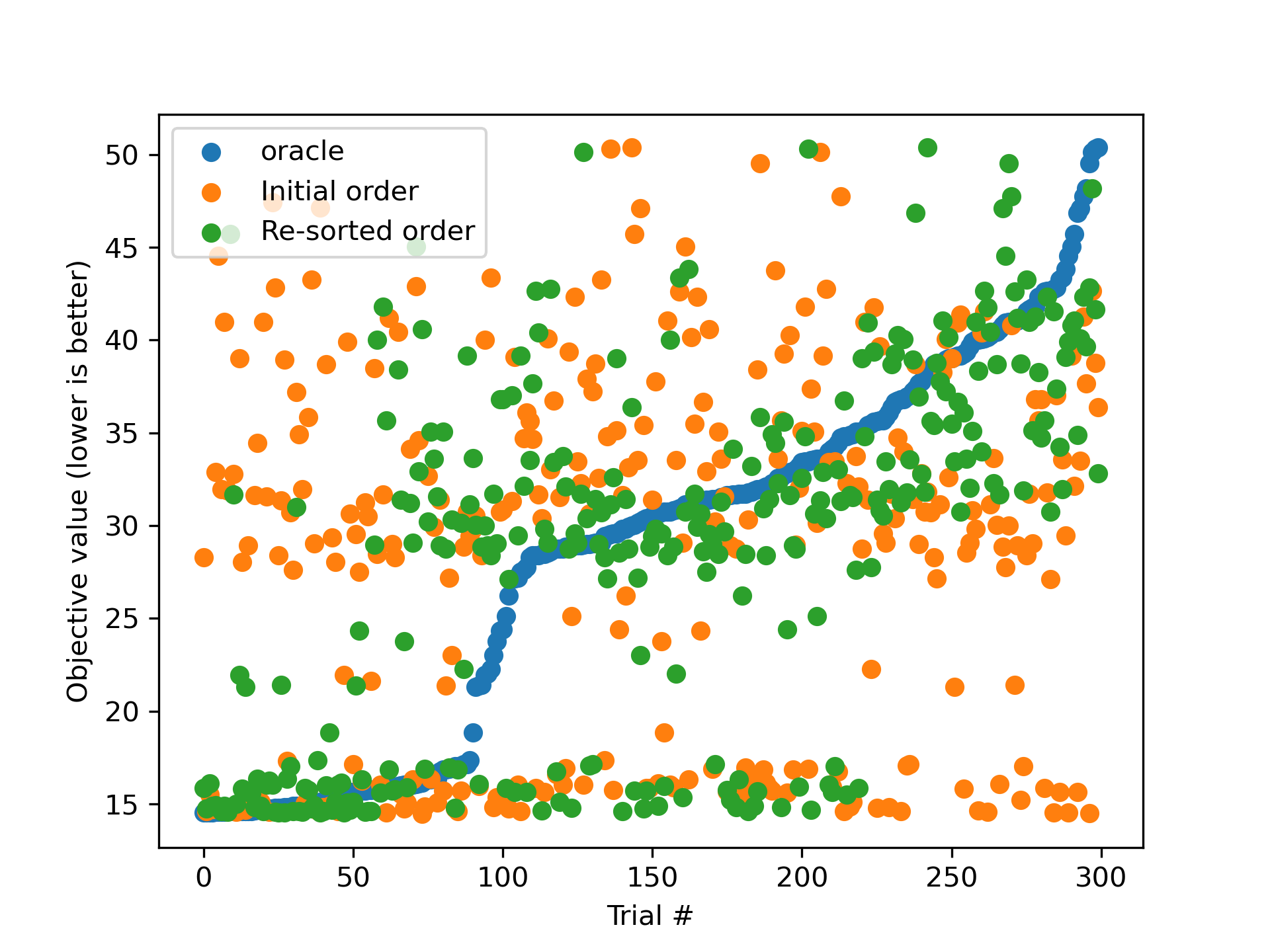}
    \caption{NCS generally moves high-quality evaluations for the Heat3D benchmark generated by GCTLA to the start of the search, improving autotuning performance.}
    \label{fig:best_NCS}
\end{figure}
\begin{figure}
    \centering
    \includegraphics[width=\linewidth]{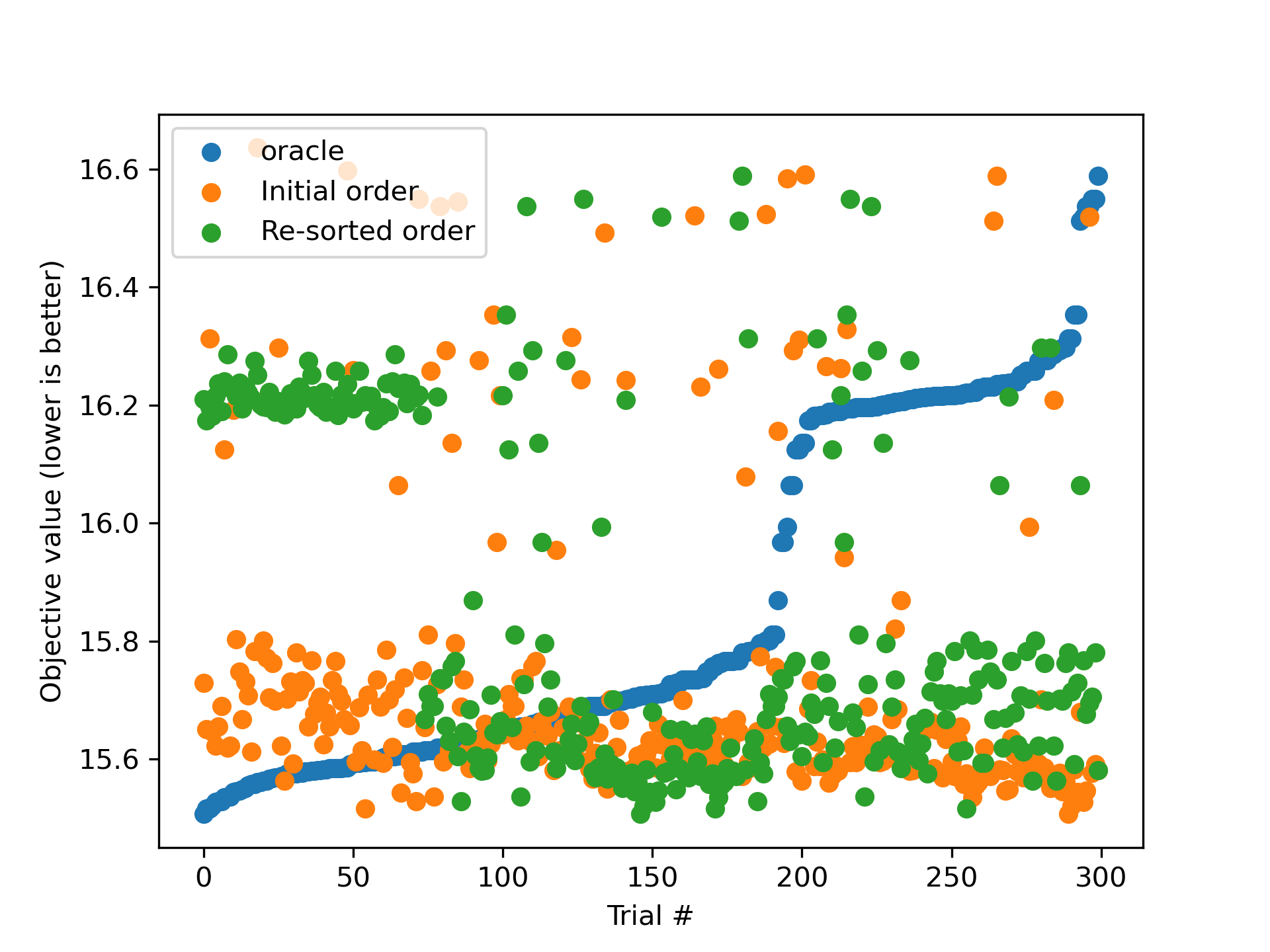}
    \caption{
    %An instance where NCS harms the GCTLA SW4Lite search by incorrectly prioritizing lower-quality results.
    NCS can harm searches by prioritizing lower-quality results.
    In this case for the SW4Lite benchmark, the majority of generations are high-quality, but NCS takes a significant number of trials to identify good performance.}
    \label{fig:worst_NCS}
\end{figure}

\tlr{Our best polybench results:
Heat3d SM (Best overall in 33 evaluations)

Our worst polybench results:
Heat3d XL (still better than other full searches in 100)

Our best ecp results:
SW4Lite XL basically tied on GCTLA as best with -100 evaluations

Our worst ecp results:
rsbench XL has bad pool from GCTLA and gets a bit worse sorted via NCS
}

As shown in Figure~\ref{fig:best_NCS}, the combination of GC+NCS grants the benefits of both techniques.
The GCTLA proposes many high-quality configurations, but due to its unordered generation, finding the best configuration is a dubious process that requires attempting each generated configuration until resources are exhausted.
With NCS, the highest-performing proposals from the GCTLA are moved to the front, generally within the first 100 evaluations, permitting a more limited autotuning expenditure to have high likelihood of extracting the best data available from the GCTLA.

However, we saw that NCS does not always improve the evaluation order.
One such example is depicted in Figure~\ref{fig:worst_NCS}, where NCS prioritizes a large number of configurations with middling performance before slowly honing in on the actual best-performing configurations.
Despite the limited data used during training, we find that a few of the better results are still available within the first-100 immediate term tuning, and the performance gap between the search-optimum and NCS's top-100 selections remains insignificant.
Unlike GCTLA, there isn't a simple manner to predict how many evaluations GC+NCS require to produce a bounded-quality result, but we see that even when NCS is not as well-suited for its task, the greatest harm it inflicts upon a larger-scale search with GCTLA is relatively minor.
Compared to the consistency of NCS dramatically improving the sorting order, we find that a successful search is more likely to be improved by NCS than degraded.

\section{Related Work} \label{sec:related} Autotuning is a well-studied problem with many different approaches suitable to different tasks.

% Topic sentence TBD: These are autotuning works that specialize in important areas that we are not considering in our work as priorities. The techniques they present should be largely orthogonal to our own work, but are not very important in our particular evaluations and should be considered as separate merits of the larger body of work

In spaces with unknown constraints such as memory availability for GPUs, tuning techniques such as BaCO~\cite{BaCO} permit dynamically learning these limits separately from the performance relationship.
Such techniques usually work best with larger data availability and longer tuning budgets than are the subjects of this work.

For real-time systems such as live databases, the variable demands imposed on the system either require separate modeling or continuous online updates.
The latter is generally preferred in literature such as Zhang et al~\cite{CloudDatabaseDynamicTuning} as it requires less data to bootstrap and continuously adapts to conditions.

Certain autotuners expose their own hyperparameters to control the search process, which may require tuning for the autotuner itself to maximize performance as discussed in Bolet et al~\cite{IPDPS_OptimizingTheOptimizer}.

Prior studies~\cite{tapus2002active, OpenTuner, wu2022autotuning, bliss} have proposed numerous techniques for autotuning HPC applications. 
A large chunk of performance gains come from highly-specific compiler optimizations, such as those seen in LLVM and GCC.
These compiler options often have to be given explicitly at compilation time or are expressed to the compiler within the source code to permit the most efficient machine-code generation.
In addition to compiler optimizations, tweaking runtime parameters and knobs in HPC applications and systems allows users to extract more performance out of such setups. 
% \tlr{Add some citations/related works demonstrating that IR performance predictions are possible}
%Innovative techniques are therefore necessary to identify such parameters that can enhance application performance.\tlr{I don't get this sentence.} \akash{better?}
% However, the ever-increasing and evolving complexity of modern HPC autotuning tasks also necessitates innovative and automated techniques for performance optimizations done at the system or application level. 

Most autotuning tasks require optimizing single- or multi-objective non-linear functions, which incentivizes widespread use of statistical techniques.
OpenTuner~\cite{OpenTuner}, for example, %is a state-of-the-art tuner that
employs various search space optimization techniques such as the simplex-based Nelder-Mead~\cite{singer2009nelder}, Torczon hill-climbers~\cite{cooper2002compilation} to minimize a single-objective HPC autotuning task faster.
More recently, Bayesian optimization has been used and adapted quite successfully to make autotuning in HPC more efficient \cite{bliss, balaprakash2018autotuning, GC_TLA, zhu2022gptuneband}.
In spite of these significant advances, most of these tools often suffer from a high code execution overhead, which adversely impacts the usability and adoption of these tools.

A commonly proposed alternative, especially for tasks targeting source code, are tools and studies that employ machine learning (ML) or artificial intelligence (AI) \cite{MGA, dutta2024mirencoder, tehranijamsaz2022learning, cummins2021programl, ben2018neural, DiffTune}.
Most of these studies achieve state-of-the-art results in a number of optimization tasks and several of them propose new code representation techniques on top of LLVM IRs, making them usable across programming languages and programming models.
However, most ML techniques must be trained specifically for each task at considerable cost, reducing their generalization capabilities while increasing the barrier to broader utilization within HPC.
% ICS reviewer was unhappy that we did not mention LLVM MCA even though it's a horrible idea: Here's what we can say about that in the space in short form
CPU simulator works such as DiffTune~\cite{DiffTune,DiffTuneRevisited} can provide cycle estimations for straight line assembly via LLVM's MCA tools, but exclude key structures \tlr{$\leftarrow$ word choice? ``features'' instead?} such as loop behaviors and function calls.
These techniques are also limited to runtime estimations only, where most autotuners can support any objective or even multi-objective tuning.
Several autotuners \cite{wu2022autotuning, wu2023ytopt, zhu2022gptuneband, GPTuneCrowd} utilize machine learning techniques such as Bayesian optimization to scale to available and relevant training data at the cost of making a number of decisions that are sub-optimal early in the search.
This increases the empirical cost of searches to enrich available information about the problem, which the advanced algorithms can use to recover the cost as time goes on with increased insight into the performance relationship.

GCTLA~\cite{GC_TLA} establishes the basic capability of probability models to represent transferrable knowledge in autotuning.
Instead of iteratively exploring and learning about the performance relationship, this technique attempts to directly sample high-performing configurations in the search space.
This approach requires far less data than ML- and AI- based autotuners, reducing the cost of specific training while still delivering high-quality results within a small number of empirical evaluations.
Another compelling characteristic of probabilistic transfer autotuning is the ability to mathematically model the number of useful evaluations to perform before collecting any evaluations in the transfer domain.
This ``budget'' can be used to determine the suitability of probabilistic sampling and the likely cost to utilize the technique prior to executing any evaluations.

\section{Conclusions} \label{sec:conclusions} \tlr{KEY CONTRIBUTIONS:
\begin{itemize}
    \item Reduced computational load and data requirement for effective long-term searches
    \item More efficient search (less evaluations to get search-local optimum)
    \item Expand capability of prior few-shot generative technique to longer-running search
    \item Permit future work on generative-predictive techniques that can benefit from this work
\end{itemize}
}

In this work, we create a framework for transfer-learning-based autotuning using LLVM IRs.
Our machine learning ensemble, NCS, is capable of prioritizing high-quality evaluations based on predicted performance from IRs.
When combined with the GCTLA probabilistic model to restrict consideration to a high-performing candidate sample, our method significantly reduces the number of necessary empirical evaluations in the search space.
Our novel combination of search space reduction with lower-fidelity discriminative modeling enables us to rapidly optimize the performance objective, outperforming existing autotuning techniques in few-shot scenarios.
We demonstrate that GC+NCS reduces the average number of evaluations needed to locate the search optimum 27.85\% of the overall search budget. \tlr{ABSTRACT / INTRO NUMBER}

NCS can be orthogonally applied to any non-iterative autotuning technique and encourages further research on probabilistic techniques and other supportive models.
Moreover, our method maintains competitive performance even in longer search durations, offering greater reach for probabilistic techniques to scale as data-efficient alternatives to traditional machine-learning-based autotuners such as Bayesian optimization and GPTune.

While our results do not indicate current failures of this kind, we know that CNNs can struggle to capture long-range dependencies or sequential relationships that exist within IR code structures, due to their localized receptive fields.
Future work may explore usage of graph neural network and transformer-based architectures to better capture global execution patterns that influence performance.

\section*{Acknowledgment}
We would like to thank NSF for their generous support in funding this project ($\#2211982$) and in part. This research was also partially supported by DOE ASCR SciDAC RAPIDS, and in part by U.S. National Science Foundation under Grants CCF-1942182. This material is based upon work supported by the U.S. Department of Energy, Office of Science, under contract number DE-AC02-06CH11357.

\bibliographystyle{ACM-Reference-Format}
\bibliography{reference} % Add BibTeX entries to reference.bib

\end{document}